\documentclass[pra,
a4paper,
showpacs,
twocolumn,
superscriptaddress]{revtex4-1}
\usepackage{amssymb}
\usepackage{amsmath}
\usepackage{amsfonts}
\usepackage{braket}
\usepackage{graphicx}
\usepackage{bm}
\usepackage{color}
\usepackage{multirow}
\usepackage{natbib}
\usepackage{hyperref}
\usepackage[normalem]{ulem}
\usepackage{verbatim}
\usepackage{dcolumn}
\usepackage{ulem}
\usepackage{float}

\usepackage[dvipsnames]{xcolor}

\def \be {\begin{equation}}
\def \ee {\end{equation}}
\def \bea {\begin{align}}
\def \eea {\end{align}}

\def \BEA {\begin{eqnarray}}
\def \EEA {\end{eqnarray}}

\def \BC {\begin{cases}}
\def \EC {\end{cases}}

\begin{document}
\title
{ 
Lateral plasmonic crystals: Tunability,  dark  modes, and weak-to-strong coupling transition    
}

\author{I.\,V.~Gorbenko}
\address{Ioffe Institute,
194021 St.~Petersburg, Russia}
\author{V.\,Yu.~Kachorovskii }
\address{Ioffe Institute,
194021 St.~Petersburg, Russia}


\begin{abstract} 
We study transmission of the terahertz radiation through a two-dimensional electron gas with a concentration controlled by grating gate electrodes. Voltage applied to these electrodes creates a lateral plasmonic crystal with a gate-tunable band structure. We find that only a part of plasmonic modes of such a crystal is seen in the transmission spectrum for the case of homogeneous excitation (so-called bright modes), while there also exist dark modes which show up only in a case of inhomogeneous excitation. We develop a theory that describes both weak- to strong- coupling transition in the crystal with increasing depth of the density modulation and a transition from resonant to super-resonant regime with increasing quality factor of the structure. We discuss very recent experiment, where  transmission of the terahertz radiation through GaN/AlGaN-based grating gate periodic structures was studied. We argue that this experiment represents an evidence of formation of the lateral plasmonic crystal with the band structure fully controlled by the gate electrodes, in a full agreement with developed theory.

\end{abstract}

\maketitle

\section{Introduction}
Active   study of  plasma oscillations in  two-dimensional (2D) electronic systems
began several decades ago \cite{chap1972,allen1977,Theis1977a, Tsui1978, Theis1978, Theis1980, Tsui1980a, Kotthaus1988} (for review, see  \cite{Maier2007}).   
Apart  from  the fundamental importance of 2D plasma phenomena, they have enormous potential for various applications, in particular,  for terahertz (THz) electronics based on high-quality  gated and ungated 2D semiconductor-based  structures. 

The interest to 2D plasmonics  increased dramatically  after Dyakonov and Shur predicted 
\cite{Dyakonov1993}
 that   a direct current (dc) in the channel of a  single-gate field effect transistor (FET)
might become unstable  with respect  to  generation of  gate-tunable  plasma oscillations in the THz range of frequencies (due to high plasma wave velocity $>10^8$ cm/s and small device size $\sim 100-1000$ nm).  The
non-linear properties of the  plasma waves in a FET channel  can be  also  used for detection of THz radiation \cite{Dyakonov1996}.  
Importantly, the coupling of the radiation with the transistor channel is enhanced  near  gate-tunable plasmonic  resonances. The quality factor of the resonances 
can be increased by using  2D structures with high mobility, particularly based on the novel materials, so that THz plasmonic devices   fabricated from graphene and topological insulators were in focus of the  research in the last decade \cite{Grigorenko2012, Vicarelli2012,DiPietro2013,Kachorovskii2013,Elkhatib2011,Giorgianni2016,Rumyantsev2015,Autore2017,Politano2017,Yang2018,Bandurin2018,Boubanga-Tombet2021,Otsuji2022}.

 The most promising way to further improve coupling  of 2D plasmons   with external radiation
 is to use systems where this coupling is  artificially increased,  such as grating  metal couplers, 2D periodic grating-gate and double-grating-gate structures. 
Metal grating couplers were used in the seminar works of Refs. \cite{allen1977,Theis1977a} to excite plasmons. However, multi-gate structures, primarily systems with a double grating gates (DGG), are much more promising. The main advantage of such structures is the possibility of periodic spatial modulation of the electron density with an amplitude controlled by  gate voltages. This makes it possible to obtain tunable plasmon resonances 
\cite{Kotthaus1988}. The    quality factors of such tunable resonances can be sufficiently high, as was shown much later by Muravjov et al. \cite{Muravjov2010} using two-dimensional multi-gate lateral superlattices based on GaN/AlGaN structures.   
This observation, supplemented by a number of theoretical and numerical studies \cite{Zheng1990, Matov1998, Mikhailov1998, Matov2002, Fateev2010, Kachorovskii2012, Popov2015, Petrov2017, fateev2019, fateev20199, Morozov2021}, triggered an  increased interest in    tunable  grating-gate  plasmonic crystals.
 
 As was understood in the last decade, the plasmonic  multi-gate structures   are  very attractive from the point of view of  possible dc-current-induced generation of plasma waves \cite{Kachorovskii2012,Koseki2016,fateev2019,fateev20199,Li2019,Kurita2014,Boubanga-Tombet2014,Bellucci2016,Petrov2017,Pan2017,Yadav2018}, and   in context of  conversion   of THz radiation into dc current  due to the so-called “ratchet” effect \cite{Ivchenko2011} existing in structures with asymmetrical spatial modulation.  The latter  effect   was actively discussed both in zero magnetic field (see \cite{Popov2011,Rozhansky2015,Faltermeier2017,Faltermeier2018,Hubmann2020,Sai2021,Monch2022,Monch2023} and a review of key publications there)  and  for sufficiently weak magnetic field (see \cite{Faltermeier2017,Faltermeier2018,Hubmann2020,Sai2021,Monch2022,Monch2023} and references therein). The modulation asymmetry can also result in the appearance of radiation-induced   traveling directional plasmons \cite{Popov2015, fateev2019, Morozov2021}.

Despite the large number of publications devoted to the multi-gate systems, a number of key issues have still not been explored in detail.  The most interesting direction for further research seems to be the use of multi-gate systems as {\it tunable plasmonic crystals} (PC). Indeed, since the charge density in the channel can be periodically modulated by gates, the plasma velocity is also modulated, leading to the formation of  pass- and stop-bands for plasma waves. It is worth noting the difference between this PC and much more studied photonic crystals: the typical wavelength of THz radiation (of the order of 100 $\mu$m) is much longer than the lattice period of a typical PC (of the order of several microns). Therefore, a crystal band structure  appears  for plasma waves, and not for electromagnetic ones. We also stress that periodic density modulation exists in the plane, so that its more natural to call such crystals as {\it lateral} plasmonic crystals. Since in the multi-gated structures the density is modulated in one direction (say, $x-$direction), the bands of the crystal corresponds to propagation of  1D plasma waves. 

The idea of a lateral tunable PC was put forward more than ten years ago \cite{Kachorovskii2012} in context of current-induced plasma wave instability and  was discussed later theoretically  
\cite{Petrov2017,Aizin2023}.  A convincing  experimental confirmation of this idea  appeared only  recently \cite{Sai2023}. In the latter work, the tunable plasmonic resonances in transmission coefficient through GaN/AlGaN-based PC  were measured. By changing  the voltage on the grating gate the 2D liquid was depleted under the gated strips, 
and appearance of
spatially isolated  plasmonic  resonances   in  highly conductive ungated regions   was demonstrated.  Although similar tunable resonances  were obtained much earlier
\cite{Kotthaus1988} and   discussed in a later numerical study \cite{Matov2002}, the results obtained in  Ref.\cite{Sai2023} were directly related by the authors   with   tunable   band structure of the lateral  PC. Specifically, it was argued  that PC  was formed in the system and was driven by changing the gate voltage  from the weak coupling regime (weak periodic density modulation in 2D electron channel) to the strong coupling regime (almost isolated stripes with high conductivity weakly connected via almost depleted regions). It was demonstrated that observed evolution of the  plasmonic resonances with gate voltage encodes information about this transition.   

Currently, there is  no consistent   theory describing    the transition between the strong and weak coupling regimes within a simple analytically solvable  model, although both regimes were discussed in literature.  In particular, all analytical calculations of the ratchet effect were performed using perturbation approach implying  weak coupling regime \cite{Rozhansky2015,Faltermeier2017,Faltermeier2018,Hubmann2020,Sai2021,Monch2022,Monch2023}. 
On the other  hand, recent  theoretical analysis of experimentally observed  dc-induced THz amplification was done in the opposite regime of a very strong coupling  \cite{Boubanga-Tombet2020}.

In this work,  we discuss different regimes of optical excitation of  lateral PC  and calculate the  transmission coefficient of external radiation throw such a crystal.   
To describe the PC, we use hydrodynamic approach, thus assuming that the  electron-electron collisions dominate over impurity  and  phonon scattering. We develop  a  theory that allows us to trace the transition from the weak coupling regime to the strong coupling regime and qualitatively explain the results of the experiment  \cite{Sai2023}.  We use a simplest model of lateral  PC  by  considering periodically repeating cells   divided into regions with different plasma wave velocities $s_1$ and $s_2.$   

We demonstrate  that  our simple model predict two types of plasmonic resonant modes---{\it bright}  and  {\it dark} modes---in agreement  with numerical analysis \cite{fateev2019}. 
The   latter ones   do  not show up  in the transmission spectrum for  symmetrical spatial modulation.   We describe modes of both types and find  conditions for observation of the  dark modes. 

The unavoidable  property of the plasmonic crystal  is  dissipation, which leads to decay of the plasmonic modes with a certain rate $\gamma$  and limits the quality factor of the resonant excitation.  We find that some modes, which  have  high frequencies  and show good  plasmonic resonances  even  for relatively large $\gamma,$  can split into number of narrow peaks with decreasing  of $\gamma.$ We call such regime of very small $\gamma$ as {\it super-resonant} regime. We also  construct   a general diagram  illustrating  weak-to-strong coupling transition with decreasing  depth of the density modulation in the channel (i.e. decreasing $s_2/s_1 $)   and resonant-to-super-resonant transition with decreasing  $\gamma.$          

We  find that not only  density modulation (which we call {\it electrical modulation}), but also the modulation of intensity of the incoming radiation ({\it optical modulation}) can change the response  dramatically. In particular,  dark modes can show up due to the radiation field modulation.   

We also modify developed results for the case of non-zero magnetic field, $B$, and demonstrate that  both band widths and distance between bands decreases  with increasing $B.$

Finally, at the end of the paper, we discuss relation of our theory   to very recent experiment   \cite{Sai2023} and find a good qualitative agreement. 
  \color{black}    
\section{Model}
\label{Sec-approach}
\subsection{Problem formulation    and general approach}

The simplest model of the lateral plasmonic crystal has been introduced in Ref.\cite{Kachorovskii2012}. This model describes a 2D electron liquid with two grating gates. By applying  two independent voltages to these gates, one  gets a system of alternating stripes  with different electron concentrations and, consequently, with  different plasma wave  velocities, $s_1$ and $s_2.$ (see also discussion of different geometries of GGS in Ref.~\cite{Shur2021}). To be specific, we assume  that $s_1>s_2$. Following Ref.~\cite{Boubanga-Tombet2020} we refer region ``1'' with higher velocity as {\it active} and region ``2'' as {\it passive} (the meaning of terms  ``active''  and  ``passive'' will be explained below).
The velocity  $s_2$  can be tuned by gate  in  the full range from $0$ to $s_1$.
If $s_1-s_2 \ll s_1,$ the plasma wave propagates in an almost homogeneous system, having a weak scattering at the boundaries between regions "1" and "2".   We    refer this case as {\it weak coupling} regime. The opposite case corresponds to complete depletion of the region "2", i.e. zero plasmon velocity, $s_2=0.$ In this case, the system is divided into a set of well-conducting strips with plasma velocity $s_1,$ separated by insulating regions, so that plasma oscillations in different conducting strips are disconnected.  Below, we  refer this case  as the {\it strong coupling} regime.
In contrast to conventional crystals,  plasmonic crystals are strictly speaking  unstable because of plasmons decay.  However,  in   the clean ballistic systems the plasmon lifetime can be sufficiently long and band structure of the PC can show up in the experiment. 

The system  is excited  by normally incident THz radiation with a wavelength exceeding the size of the crystal cell  $\lambda \gg L_{1}+L_{2}$ (this condition allows to avoid grating-induced diffraction orders). Light is linearly polarized along $x$-direction (perpendicular to the grating stripes).  We will calculate  the transmission coefficient  $T$ in the presence of external magnetic field perpendicular to the plasmonic superlattice.

    Radiation excites plasmonic oscillations in the modulated 2D channel.    We assume that conductivity of the electron liquid in the  2D channel is much smaller than the speed of light $2\pi \sigma/c \ll 1.$  In this case, one can neglect radiative decay of plasmonic oscillations (see Refs.~\cite{Mikhailov1998, Boubanga-Tombet2020}) and 
     transmission coefficient is close  to unity with a small correction, which  
        is  fully expressed in terms of ohmic dissipation in the channel, $P$:  
\be T \approx 1-\frac{8 \pi \, P}{c \sqrt{\epsilon} E_0^2 },
\label{T}
\ee 
where $E_0$ is the amplitude of the incoming radiation,   and 
  $\epsilon$ is the dielectric constant, which, for simplicity,  is assumed to be the same everywhere.  
Actually, this correction is the first dominating term  in the expansion of $(1-T)$ over $1/c.$   
  Hence, in order to find $T,$ we  need to calculate radiation-induced dissipation in the channel $P.$ The latter    encodes information  about  the plasmonic resonances.     
  
To describe plasma modes, we assume that electron-electron collisions are very fast as compared to the momentum relaxation rate. This  allows us to use standard hydrodynamic approximation.     
  Local dissipation within this approximation is given by $ m |\mathbf v (x,t) |^2/\tau$ \cite{Rozhansky2015},  where 
$\mathbf v (x,t)$ is the hydrodynamic velocity and $\tau$ is the momentum relaxation  time.
Hence, the         
ohmic dissipation per unit area in a plasmonic crystal reads 
\begin{equation}
    P = \,  \left< N \frac{m |\mathbf v (x,t) |^2}{\tau} \right>_{x,t}
\label{Eq-diss0}
\end{equation}
with $N=N(x,t)$ as the electron concentration in the channel and   $\left<\dots\right>_x$ as averaging taken over area of the crystal.

\subsection{External radiation} 

Next, we discuss the properties of the external field in the  2D channel.    
Importantly, not only the charge density is modulated in the 2D  channel due to the grating gates, but also the intensity of the electromagnetic wave.  Hence, there are {\it electrical} and  {\it optical} modulation. Accordingly, the  electric field  in the channel can be represented as the sum of a homogeneous component and a component {\it optically} modulated with the period of the superlattice. Both components oscillate with a THz frequency $\omega.$

Rigorous calculation of the modulated field is a tricky problem which implies solution of 3D electromagnetic equations. Such analysis is out of scope of the current paper.  Some approximations for solution  of such  3D problem can be found in  Ref.~\cite{Mikhailov1998}. 

Here we will limit ourselves to a simple approach to describe field modulation by a grating, proposed in Ref. \cite{Ivchenko2011}. This method was  successfully used for explanation of number of experiments on  photovoltaic of the grating gate structures Refs. \cite{Rozhansky2015,Faltermeier2017,Faltermeier2018,Hubmann2020,Sai2021,Monch2022,Monch2023} .   
The most general form of modulated field  in 2D channel reads $E(x,t) = E_0 [1+ \sum_{n=1}^{\infty} h_n \cos(n k  x + \phi_n)] \cos \omega t,$ where $E_0$ is the homogeneous amplitude of the incoming wave,
\begin{equation}
k=\frac{2\pi}{L_1+L_2}
\label{k-vector}
\end{equation}
is reciprocal lattice vector, $h_n>0$ and $\phi_n$ are the modulation strength and the phase of $n-$th harmonic, respectively.  Following Ref.~\cite{Ivchenko2011}, we assume  $ h_{n\neq 1}  \ll h_1$ and keep the first harmonic only:
\begin{equation}
 E(x,t) \approx  E_0[1+  h \cos(k x + \phi)]\cos \omega t .  
 \label{Eq-grating_modulation}
\end{equation}
Here, we put $h_1=h, \phi_1=\phi.$  Hence, we assume that field is modulated and that the modulation depth is weak. It is worth noting that Eq.~\eqref{Eq-grating_modulation} describes external field, with $h$ arising due to weak optical modulation, while dynamical screening of this field by electron liquid will be discussed below and is not assumed to be weak.  
A special comment is needed concerning the phase $\phi,$  which controls asymmetry of the structure. Indeed, 
choosing the  edge of region ``1'' at the origin of the coordinate, $   x=0,$  we find that the structure  has spatial inversion  symmetry with respect to  $x_0=L_1/2$  
provided  that $cos (k x+ \phi)$ is invariant under such inversion. This is the case, when $\phi$ equals to special values
\be
\phi_{\rm sym} = \pi \left( \frac{  L_2}{L_1+L_2} +  n \right), 
\label{phi-sym}
\ee
where $n$ is integer.  In realistic structures, asymmetry can be created by fabricating of additional gratings and may also occur due to imperfections in the technological process  (see discussion in Ref.~\cite{Ivchenko2011}).     

External radiation \eqref{Eq-grating_modulation} consists of two contributions, homogeneous and inhomogeneous, both of which  excite plasmonic oscillations  in active and passive regions of the  2D channel. Importantly, the homogeneous component
\begin{equation}
 E(x,t) = E_0 \cos{(\omega t)}
\end{equation}
excites only a part of the plasmonic modes, so-called {\it bright} modes. 
At the same time, due to the inhomogeneous component, a number of other modes can be excited, the so-called {\it dark} modes provided that $\phi \neq \phi_{\rm sym} $.

Equation  \eqref{Eq-grating_modulation} has a form of a standing wave. 
Another type of modulation with non-zero in-plane momentum,     
\begin{equation}
 E(x,t) = E_0 \cos{(K x -\omega t)},
 \label{Eq-traveling_wave}
\end{equation}
appears  when   incoming radiation   has  non-zero angle of    incidence.  Such a traveling wave can be also used to probe the dark states as was recently demonstrated  for visible light scattering on meta-surface of metallic particles   \cite{Hakala2017}.

\subsection{Hydrodynamic approximation} 
\label{HD}
We  assume that electron-electron collisions dominate over impurity and phonon scattering and describe the   electron liquid  in the FET channel  by hydrodynamic  (HD) equations---the Euler equation (including  external electric field force and Lorentz force)  and  the  continuity equation:

\begin{align}
&\frac{\partial \mathbf{v}}{\partial t}\! + \!(\mathbf{v} \cdot \mathbf{\nabla}) \mathbf{v} \!+\! \gamma \mathbf{v}\!=\! - \frac{e}{m} \mathbf{\nabla} U +  \left[ \boldsymbol{\omega}_{\rm c} \times \mathbf{v} \right]+\frac{\mathbf{F}}{m},
\label{Eq-Navier_Stokes}
\\ 
&\frac{\partial U}{\partial t} + {\rm div} (U \mathbf{v}) = 0. 
\label{Eq-continuity}
\end{align}
Here $\mathbf v(x, t)$ is the local drift velocity, $U=U(x,t)$ is the local voltage swing between 2D channel and gate electrode, and $\omega_c$ is the cyclotron frequency.  Spatial derivative of this potential   
is related to  dimensionless concentration  in the channel, $$\delta n=\frac{N-N_0}{N_0},$$ as follows:
$$\frac{e}{m} \frac{\partial U}{\partial x} = s^2 \frac{\partial \delta n}{\partial x},$$
where 
\begin{equation} 
s = \sqrt{\frac{e (U_g-U_{th})}{m}} 
\label{Eq-ss}
\end{equation}
is the plasma wave velocity controlled by the gate voltage $U_g,$ and $U_{\rm th}$ is the so-called threshold voltage.  The stationary electron concentration in the channel is also controlled by $U_g$:
\begin{equation}
 N_0=\frac{C (U_{\rm g}- U_{\rm th})}{e},    
 \label{Eq-N0}
\end{equation}
where $C=\epsilon/4\pi d $ is the channel capacitance per unit area, $d$ is the spacer width, and  $\epsilon$ is the dielectric constant.   

Dissipation in the channel is controlled by momentum relaxation rate $\gamma.$ We neglect here relaxation related to the   viscosity of the electron liquid, $\eta,$ assuming that $\eta q^2 \ll \gamma$ for typical wave vectors $q.$  We also assume that there is a constant magnetic field perpendicular to the plane of the structure, which creates the Lorentz force, so that we added the term $\left[ \boldsymbol{\omega}_{\rm c} \times \mathbf{v} \right]$ to Eq.~\eqref{Eq-Navier_Stokes}.

 THz radiation with a linear polarization in direction perpendicular to the grating creates the external force  $ \mathbf F(x,t) = e  \mathbf n_x E(x, t),$ where $\mathbf n_x$ is the unit vector in $x$ direction and  $E(x, t)$ is the field amplitude which can be optically modulated as explained above. Total field acting in the channel is given by the sum of the optically modulated  external field and plasmonic force: $$\frac{eE_{\rm tot}}{m} =\frac{eE}{m} - s^2 \frac{\partial \delta n}{\partial x}.$$

 We assume that   stripes   with plasma wave velocities $s_1$ and  $s_2$  have corresponding  lengths  $L_1,$ and  $L_2,$ respectively. In order to solve Eq.~\eqref{Eq-Navier_Stokes} and \eqref{Eq-continuity}, we  use standard   boundary conditions between regions \cite{Kachorovskii2012, Petrov2017} that correspond to the current  and energy flux conservation on the boundary between the strips (we assume here that dc current in the channel is absent):
\begin{equation}
    s^2 \delta n = {\rm const}, \, \,  s^2 v_x = {\rm const}.
\label{Eq-BC}
\end{equation}
Although above approximation is very simplified, in particular, because it does not take into account  the  ungated   gaps  between gated regions, which always present  in realistic structures,  it captures all basic effects  described in numerical simulations \cite{Matov2002,Fateev2010,Popov2015, fateev2019, Morozov2021}  (see also recent discussion  of the effect of ungated gaps on the plasmonic spectrum \cite{Zarezin2023}).   

 Next,  we linearize  Eqs.~(\ref{Eq-Navier_Stokes}) and   (\ref{Eq-continuity}) with respect to external radiation,   find the  radiation-induced velocity, and calculate dissipation by using  Eq.~\eqref{Eq-diss0}.   
  We delegate  some technical details to Appendixes, focusing in the main text  on key steps  only.

In the first part of the paper, we study system without optical modulation, thus assuming that the amplitude of the external field  $E$ does not depend  on $x$ (i.e. $h=0$). Generalization for  the case of inhomogeneous excitation   will be discussed below in the Section  \ref{Sec-light_modulation}.   

Assuming that   solutions of linearized Eqs.~\eqref{Eq-Navier_Stokes},\eqref{Eq-continuity}    are  proportional to  $ e^{-i \omega  t},$ we find (see Appendix \ref{AppHD})
\be
\begin{aligned}
&\delta n =A e^{i q x}+B e^{-i q x}, 
\\
&v_x=(A e^{i q x}-B e^{-i q x})\frac{\omega}{q}+ v_{\rm ext},
\\
&v_y=v_x \frac{-i \omega_c}{\omega+i \gamma}.
\end{aligned}
\label{vxy}
\ee
Here
\be
v_{\rm ext}=\frac{i F_0 (\omega+i \gamma)}{2 m \left[ (\omega+i \gamma)^2-\omega_c^2 \right]}
\ee
is the velocity due to external homogeneous excitation,  
\be
q  = \frac{\Omega+ i \Gamma}{s}
\label{Eq-wv_with_B}
\ee
is a complex wave vector,   $\Omega $ and $  \Gamma$ are real parameters found from
\be
\Omega+ i \Gamma=\left[ {\omega}\frac{(\omega+i \gamma)^2-\omega_c^2}{\omega+i \gamma} \right]^{1/2},
\label{W-G}
\ee
 and  $A=A_{M,\alpha},~B=B_{M,\alpha}$ are unknown amplitudes that depend on the number of PC cell $M$ and type of the stripe within the cell  $\alpha=1,2.$

 The wave vector $q=q_{\alpha}$  depends on $\alpha$  due to different plasma velocities,   while    $\Omega$ and $\Gamma$  are the same in both  regions if $\omega_c$ and $\gamma$ are constant across the PC. We also notice that for $\omega_c = 0,$  expression for  wave vector $q_\alpha$ simplifies: 
 \be
q_\alpha (\omega_c = 0) = \frac{\sqrt{\omega(\omega+i \gamma)}}{s_\alpha}.
\label{Eq-wv_without_B}
\ee
 
Strictly speaking, the PC  is stable only for  $\gamma=0.$ In this case, and in the absence of the external radiation, $F=0,$ the spectrum of the crystal $\omega_n(K) $ ($n$  numerates bands of PC) is found from the standard condition
\be
{\rm det}|e^{i K (L_1 +L_2)} - \hat T_1 \hat T_2 |=0,
\ee
where $K$ is quasimomentum of the crystal and $T_{1,2}$ are  transfer matrices of regions  $1$ and $2.$ Analytical expressions for these matrices are given in Appendix~\ref{AppHD}.   

For $\gamma \neq 0$ decay of the plasmonic oscillations due to momentum scattering can be compensated by the energy gain from the THz radiation. In this case, the only solution which is finite for $x \to \pm \infty$ does not depend on the cell number $M$ (for the case of the homogeneous excitation discussed in this section).  Skipping index $M$ and       
introducing vectors 
\be
\Psi_1 = 
\begin{pmatrix}
  A_1 \\ 
  B_1 
\end{pmatrix},
\Psi_2 = 
\begin{pmatrix}
  A_2 \\ 
  B_2 
\end{pmatrix}
\ee 
we arrive at  a system of coupled equation determining the  solution which is finite for $|x| \to \infty :$   
\be
\begin{aligned}
&\Psi_1 = (1-\hat{T}_2 \hat{T}_1)^{-1} (\hat{T}_2 f_1 + f_2) \textbf{e}, \\
&\Psi_2 = (1-\hat{T}_1 \hat{T}_2)^{-1} (\hat{T}_1 f_2 + f_1) \textbf{e}, 
\end{aligned}
\label{Eq-T-f}
\ee
where  $\textbf{e} = (-1,1),$  and  transfer matrices $\hat{T}_1,~ \hat{T}_2$  and   solutions  $f_1,~ f_2$ are given in Appendix~\ref{AppHD}.

Direct calculation of dissipation (see Appendix~\ref{AppHD}) yields
\be
\begin{aligned}
&P = \left( 1+ \frac{\omega_c^2}{\omega^2+\gamma^2}\right) \frac{F_0^2 C \gamma \omega^2 }{2 e^2 (L_1+L_2) (\Omega^2 + \Gamma^2)^2}
\\
& \times \left[ (L_1 s_1^2 + L_2 s_2^2)   + \frac{(s_1^2 - s_2^2)^2  \rm{Re}\left[ (\Gamma-i \Omega )^3 \Sigma  \right] }{  \Omega \Gamma (\Gamma^2 + \Omega^2) |\Sigma|^2}  \right]
\end{aligned}
\label{Eq-maindis}
\ee
with
\begin{equation}
    \Sigma = s_1 \cot{q_1 L_1/2}+s_2 \cot{q_2 L_2/2}.
    \label{sigma}
\end{equation}
Dissipation $P$ has maximum   when $|\Sigma|^2$ has minimum. We notice that for $\gamma=0,$  $\Sigma$ is real and has exact zeros at frequencies that determine plasmonic resonances.

Expression for dissipation Eq.~(\ref{Eq-maindis}) is the main analytical result of this section. It  represents all possible regimes of PC for arbitrary damping rate, magnetic field  and  $s_2/s_1$ ratio.

\section{Spectrum of plasma waves: bright and dark modes. Strong and weak coupling ($B= 0$).   } 
\label{sec-spectrum} 
\subsection{Bright and dark modes} \label{bright-dark}
 For zero magnetic field, $\omega_{\rm c}=0,$  and in the absence of the  momentum relaxation, $\gamma=0,$
 spectrum of  the lateral PC, $\omega (K), $  obeys \cite{Kachorovskii2012}:
\be
\begin{aligned}
&\cos{\left[K(L_1+L_2)\right]}
\\
& = \cos{\frac{\omega L_1}{s_1}}\cos{\frac{\omega L_2}{s_2}}-\frac{s_1^2+s_2^2}{2 s_1 s_2} \sin{\frac{\omega L_1}{s_1}} \sin{\frac{\omega L_2}{s_2}}.
\end{aligned}
\label{Eq-natural_frequency}
\ee

Solutions of this equation for $K=0$ are of particular interest,  because, at first glance,   all these modes could be excited by normal incident light with zero in-plane momentum and $h=0.$   It turns out, however, that only half of frequencies found from  Eq.~\eqref{Eq-natural_frequency} show up
in the excitation spectrum, so that there are so-called bright and dark modes    (numerical analysis of these modes in the grating gate structures is presented in Ref.~\cite{fateev2019}).

Next, we discuss this issue in more detail.
For $K=0$ one can rewrite  Eq.~\eqref{Eq-natural_frequency} as a product of two frequency-dependent terms:  
\begin{equation}
Q_{\rm bright} (\omega) Q_{\rm dark} (\omega)=0,
\end{equation}
where
\begin{align}
&Q_{\rm bright} = s_1 \cos{\frac{\omega L_1}{2 s_1}} \sin{\frac{\omega L_2}{2 s_2}} + s_2 \cos{\frac{\omega L_2}{2 s_2}} \sin{\frac{\omega L_1}{2 s_1}} ,
\label{Eq-Qbright}
\\
 & Q_{\rm dark}  = s_2 \cos{\frac{\omega L_1}{2 s_1}} \sin{\frac{\omega L_2}{2 s_2}} \! + \! s_1 \cos{\frac{\omega L_2}{2 s_2}} \sin{\frac{\omega L_1}{2 s_1}}  .
\label{Eq-brackets}
\end{align}

Hence, for $K=0$ there are two sets of solutions found from conditions   $Q_{\rm bright}(\omega)=0$  and  $Q_{\rm dark}(\omega)=0,$ respectively.   Comparing  Eq.~\eqref{sigma} with Eqs.~\eqref{Eq-Qbright}  and  \eqref{Eq-brackets}, we see that   frequencies obeying the condition  $Q_{\rm bright}=0$  also obey  $\Sigma=0$ (at $\gamma=0$) thus providing resonance in $P$ in contrast to  solutions of condition   $Q_{\rm  dark}=0,$  
which do not show up in Eq.~\eqref{Eq-maindis}.

 Next we  consider two different regimes  of  coupling   between  the crystal and the external radiation.

\subsection{Strong coupling}

 In the case $s_1 \gg s_2,$  there are two series of the resonant frequencies   found from Eq.~\eqref{Eq-Qbright}: 
  \be
  \omega_1^n=(2 n +1) \omega_1, \qquad \omega_2^n=2 n  \omega_2,  
   \ee
  where 
  \be
  \omega_1= s_1 k_1, \qquad \omega_2= s_2 k_2
  \label{omega12}
  \ee
 are the fundamental frequencies in the  regions ``1'' and ``2'', respectively,  
 and 
 \begin{equation}
     k_{1,2} =\frac{\pi}{L_{1,2}}.
 \end{equation}
 We notice that  $\omega_2 \ll \omega_1$ for  $L_1 \sim L_2.$  Hence,     if there is any type of broadening of the plasmonic resonances, say, because of  the finite $\gamma,$  such that $\omega_2 \ll \gamma \ll \omega_1,$ the resonances in the region  ``2'' overlap forming continuous spectrum of plasma excitation, while  sharp resonances in the region ``1'' survive.  In this regime, we have independent   stripes of the type ``1'', separated by dissipative  regions of type ``2'' \cite{Boubanga-Tombet2020}.  Hence, following \cite{Boubanga-Tombet2020}   we call stripes  of type ``1'' as {\it active}, while stripes  of type ``2'' as {\it passive}.

\subsection{Weak coupling}

The weak coupling regime corresponds to weak modulation of the electron density, i.e. small difference in plasma wave velocities in the  neighboring  regions, $$\frac{s_1-s_2}{s_1} \ll 1.$$
In this case,   the  only relevant wave-vector in the problem is  $k$  [see  Eq.~\eqref{k-vector}]
and the resonant plasmonic frequencies found  as solutions of $Q_{\rm bright}=0$ are given by
\begin{equation}
    \omega_n^{\rm weak} \approx  n s_1 k,    
     \label{Eq-w_weak}
\end{equation}
where $n \neq 0$ is  the integer number (the case $n=0$ corresponds to the Drude peak).

\begin{figure}[h!]
\centering
\includegraphics[width=8.6 cm]{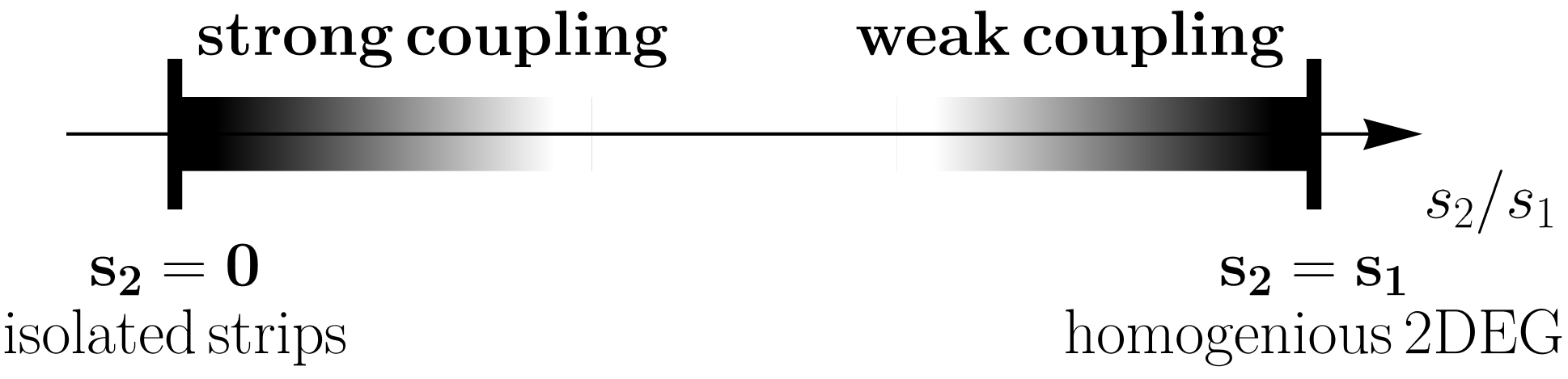}
\caption{Different regimes of  plasmonic oscillations in  non-dissipative  PC ($\gamma=0$):  (i) weak coupling regime,  $s_1 \approx s_2,$     corresponding  to  the weak modulation of the density in the 2D channel [the positions of resonances are given by Eq.~\eqref{Eq-w_weak}]; (ii) strong coupling regime, $s_2 \to 0,$ corresponding to a very strong modulation, i.e. isolated strips ``1''  with high concentration separated by   almost  fully depleted  strips ``2'' (resonances are well described by $\omega \approx (2n+1) \omega_1$ in the  active region   and $\omega \approx 2 n \omega_2$  in the passive region).     
}
\label{Fig-regimes_weak_strong}
\end{figure}

\begin{figure}[h!]
\centering
\includegraphics[width=8.6 cm]{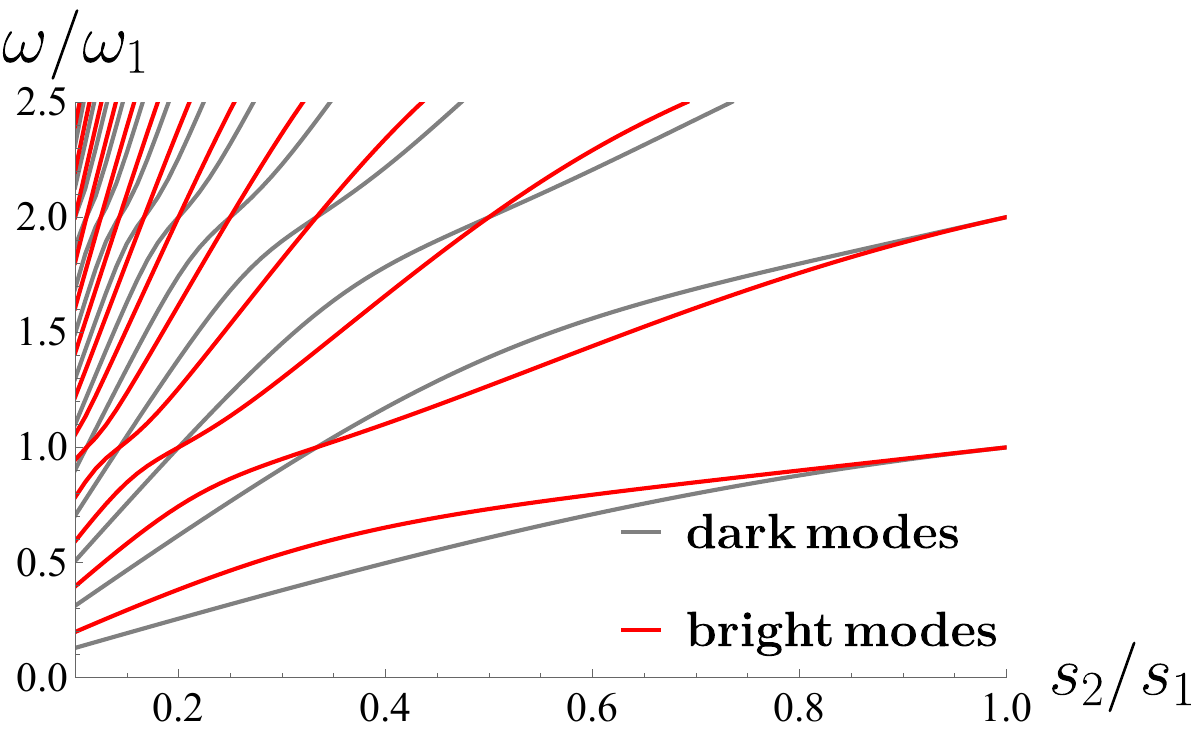}
\includegraphics[width=8.6 cm]{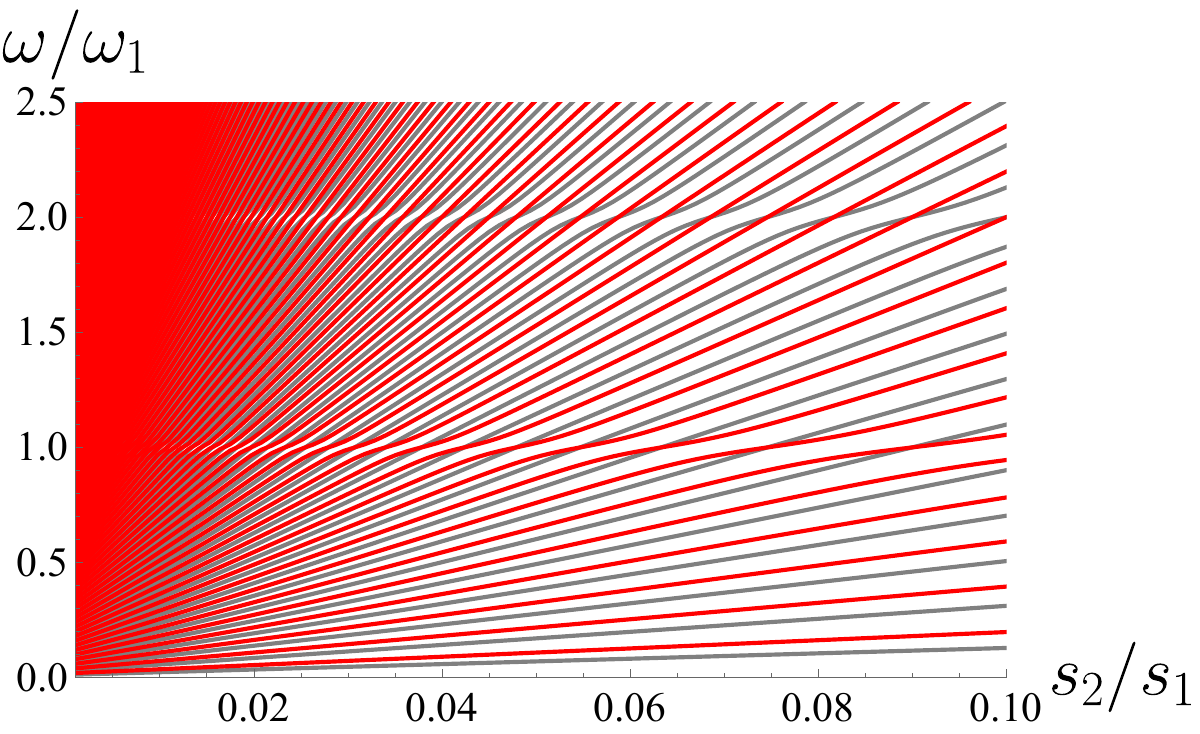}
\caption{Dependence of the frequencies $\omega_n^{\rm bright}$ (red lines)   and  $\omega_n^{\rm dark}$ (grey lines)  on the ratio $s_2/s_1$ for $L_1 = L_2.$  Upper panel shows interval $0.1<s_2/s_1<1,$ lower panel shows interval  $0<s_2/s_1<0.1.$  As seen from the lower panel, spectrum of resonances becomes infinitely dense in the limit  $s_2 \to 0$ because distance between neighboring levels  turns to zero:   $\Delta \omega  \sim s_2/L_2 \to 0 $.  The lower panel  does  not show any  structure  in the limit $s_2 \to 0$ in contrast to the dissipation map (see Fig. \ref{Fig-resfreq} below)}
\label{Fig-natfreq}
\end{figure}

\subsection{Weak-to-strong coupling transition}
\label{Sec-eigen_freq}
Solutions $\omega_n^{\rm bright}$    and  $\omega_n^{\rm dark}$  found from $Q_{\rm bright} (\omega)=0$ and  $Q_{\rm  dark}(\omega)=0,$ respectively,  depend on the ratio $s_2/s_1.$  These dependencies are plotted in  Fig.~\ref{Fig-natfreq} in the interval $ 0.1< s_2/s_1<1$ [panel (a)] and separately with the use of different interval  $ 0< s_2/s_1<0.1$ [panel (b)]. Similar dependencies of plasmonic frequencies on electron concentration can be found in Ref.~\cite{Aizin2023}. 

 For $s_2 \to 0,$   Eq.~\eqref{Eq-natural_frequency} yields  
\begin{equation}
   0= \sin{\frac{\omega L_1}{s_1}} \sin{\frac{\omega L_2}{s_2}}=  4 Q_{\rm bright} Q_{\rm dark}/s_1^2 
\label{s2to0}
\end{equation}
with 
\begin{align}
    &Q_{\rm bright} \approx s_1  \cos{\frac{\omega L_1}{2 s_1}} \sin{\frac{\omega L_2}{2 s_2}}  
    , \label{Qb} \\
    &Q_{\rm dark} \approx  s_1   \sin{\frac{\omega L_1}{2 s_1}} \cos{\frac{\omega L_2}{2 s_2}} ,
    \label{Qd}
    \end{align}
in this limit. As seen from  Eq.~\eqref{Qb}, distance between neighboring levels is proportional to  $\Delta \omega   \sim s_2/L_2$ and decreases with  decreasing $s_2.$ As we demonstrate below in Sec.\ref{Sec-regimes}, corresponding resonances overlap and disappear    in the    total dissipation $P$ provided that  $\Delta \omega$ becomes smaller than the dissipation rate $\gamma.$ However, resonances  $\omega_m^{\rm strong}$ survive and the system still show resonant behavior   (blue region  in  Fig.~\ref{Fig-regimes}).

From Fig.~\ref{Fig-natfreq} we see, that the  frequencies of bright and dark modes cross each other for certain values of $s_2/s_1.$  
This happens for two discrete  series of $s_2/s_1,$ found from  the condition $Q_{\rm bright}(\omega)=Q_{\rm dark}(\omega)=0$: 
\begin{equation}
   \frac{s_2}{s_1} = \frac{L_2}{L_1} \frac{1+2 n}{1+ 2 m}, \quad {\rm with} \quad \omega = (2 n+1) \frac{\pi s_1 }{L_1},
\label{Eq-cond1}
\end{equation}
and 
\begin{equation}
    \frac{s_2}{s_1} = \frac{L_2}{L_1} \frac{ n}{ m}, \quad {\rm with} \quad  \omega = 2 n \frac{\pi s_1}{L_1}.   
\label{Eq-cond2}
\end{equation}
Here  $n$ and $m$ are integer. It is worth noting that frequencies of the intersections depend on $n$ only,  so that solutions for fixed $n$ and different $m$ belong to  the same horizontal line (see Fig.~\ref{Fig-natfreq}).    
Also solutions cross at $s_2 = s_1$  for $\omega = 2 \pi s_1/(L_1+L_2)$ [in this trivial case  $Q_{\rm dark} (\omega) \equiv Q_{\rm bright} (\omega)$]. 


\section{Dissipative PC regimes in the absence  of optical modulation    }
\label{Sec-regimes}
In the previous section we discussed non-dissipative case $\gamma = 0.$ 
Let us now assume $\gamma \neq 0$ and discuss dissipation in PC at zero magnetic field. In this section, we focus on the simplest case assuming that optical modulation is absent, $h=0.$   

\begin{figure}[h!]
\centering
\includegraphics[width=8.6 cm]{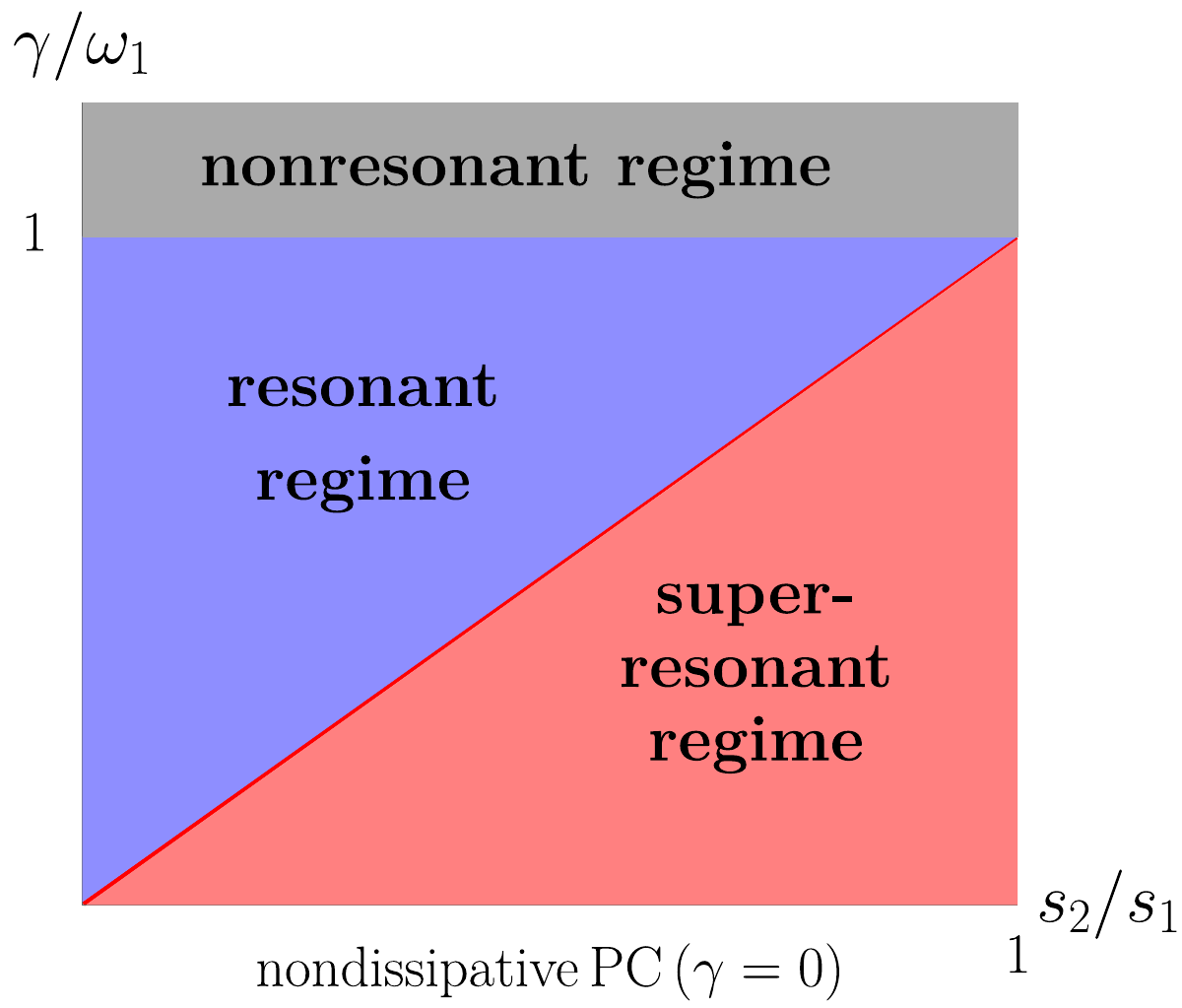}
\caption{Different  regimes of  excitation of the   PC   by  homogeneous external field at $L_1=L_2.$ For $\gamma=0$ the PC is non-dissipative. For small $\gamma$ ($\gamma \ll \omega_2$), the dissipation is weak and the response of the PC  demonstrates  sharp  {\it super-resonances} at  frequencies of  bright modes (red region). With increasing $\gamma$ above $\omega_2$    super-resonances  overlap and the response shows {\it resonances} at bright frequencies   in the active region  (blue region). With increasing $\gamma$ above $\omega_1,$ all resonances are suppressed  and we fall into the  non-resonant regime (grey region).   }
\label{Fig-regimes}
\end{figure}
The problem is different as compared to a  single plasmonic resonator with a  single fundamental frequency $\omega_0$  say single FET, which shows  only two excitation regimes -- {\it resonant} ($\gamma \ll \omega_0$) and {\it non-resonant} ($\gamma \gg \omega_0 $).  

By contrast, in a PC there are two characteristic fundamental frequencies, $\omega_1$ and $\omega_2.$ In the strong coupling case, when these frequencies are essentially different ($\omega_1  \gg \omega_2)$ [see Eq.~\eqref{omega12}], there are three different  regimes: non-resonant regime of very high damping, $\gamma \gg \omega_1$;  resonant one corresponding to overlapping  the plasmonic resonances in the passive regions,   $ \omega_2\ll \gamma \ll \omega_1$;   and   also  the regime that can be realized in the high-quality structures,     $  \gamma \ll \omega_2 \ll \omega_1.$  We call the latter case {\it super-resonant} regime.   

\begin{figure}[h!]
\centering
\includegraphics[width=8.6 cm]{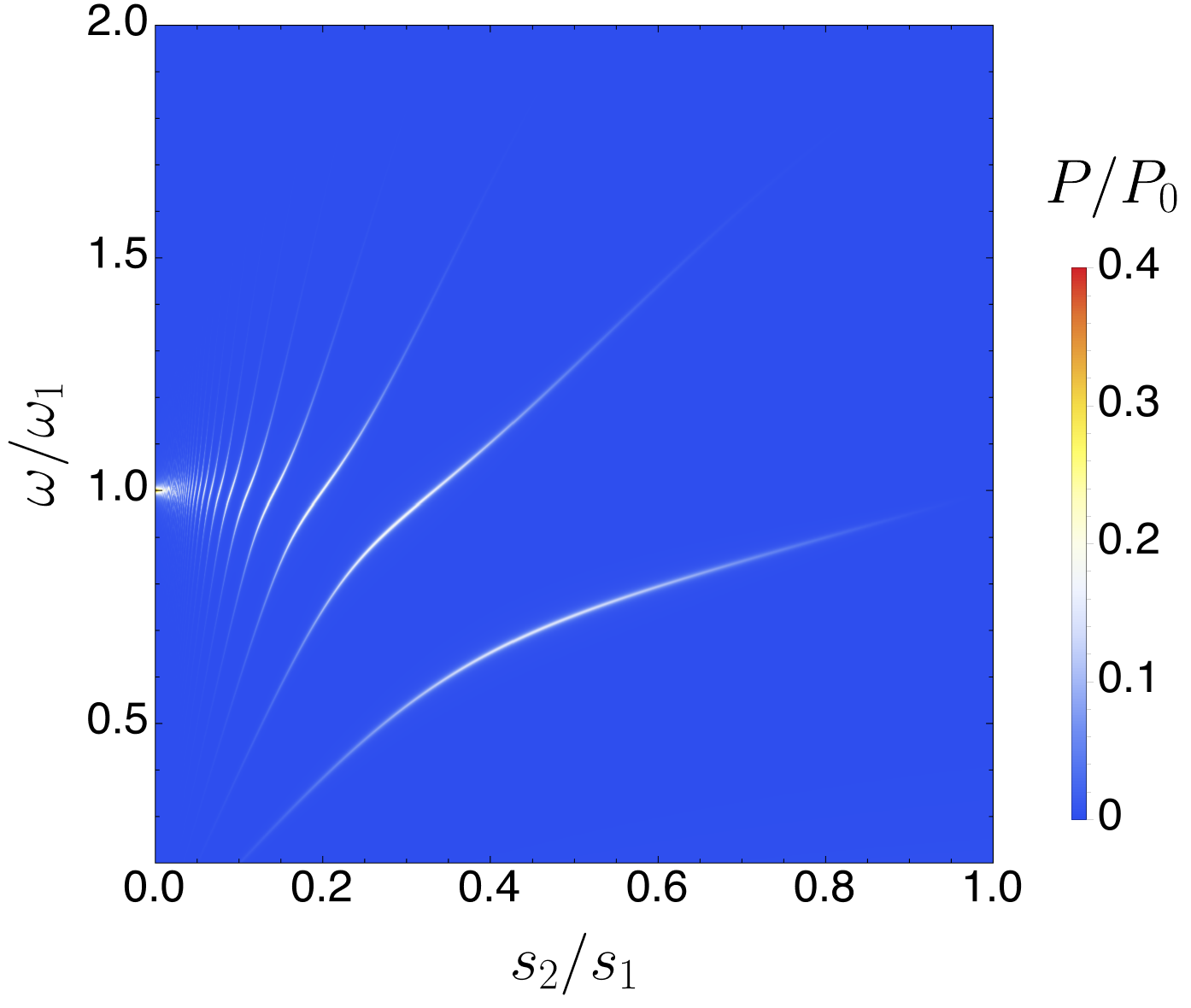}
\includegraphics[width=8.6 cm]{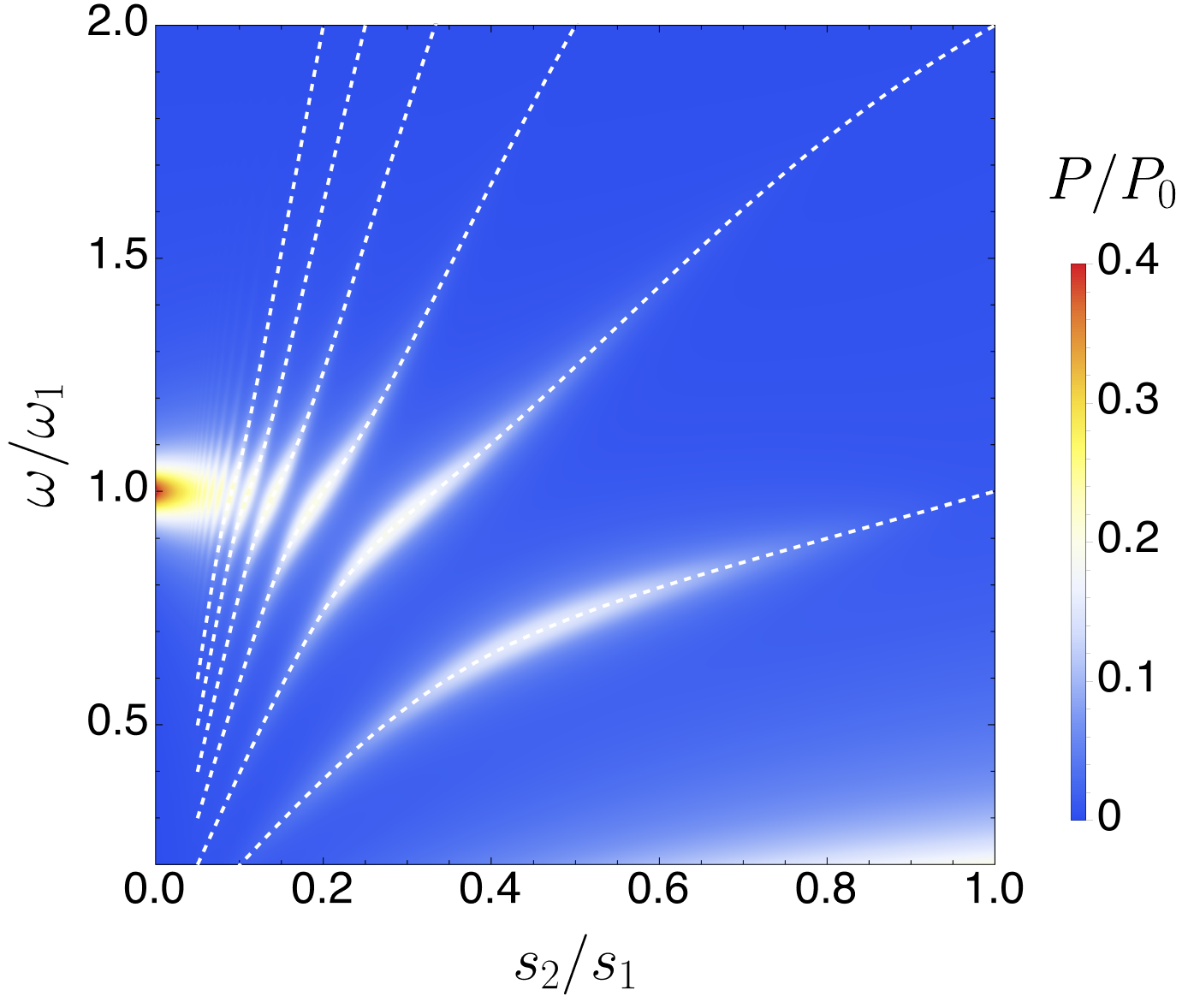}
\caption{Dissipation map  in the plane $(\omega/\omega_1,s_2/s_1)$   for $L_1=L_2$ and different damping rates:   $\gamma = 0.01 \omega_1$ (upper panel) and  $\gamma = 0.1 \omega_1$ (lower panel). 
Cross-section of upper panel at $s_2/s_1 = 0.05$ corresponds to super-resonant
regime, while cross-section of lower panel at the same ratio of $s_2/s_1$ corresponds to the resonant regime. 
The $\omega/\omega_1$-axis is limited from below with the value $0.2$ 
to hide high-amplitude Drude peak (the Drude peak   is shown  in Fig.~\ref{Fig-evolution}). }
\label{Fig-resfreq}
\end{figure}

It is convenient  to illustrate    these  regimes by using diagram  in the plane  $(s_2/s_1,\gamma/\omega_1)$.  Schematically, boundaries of these excitation regimes  are defined by conditions $\gamma = \omega_2 $, $\gamma = \omega_1$ and $s_2 =s_1.$   Corresponding straight  lines separate   different regions in Fig.~\ref{Fig-regimes}.
Typical  maps of dissipation  corresponding to different values of $\gamma$ are shown in Fig.~\ref{Fig-resfreq}. Cross-sections of these maps by vertical lines correspond to different areas in  Fig.~\ref{Fig-regimes}. For example, cross-section for upper panel of Fig.~\ref{Fig-resfreq} at $s_2/s_1 = 0.05$  corresponds to super-resonant regime, while  cross-section for lower panel of Fig.~\ref{Fig-resfreq}b at the same ratio $s_2/s_1$   corresponds to the resonant regime.    
 Below we  derive analytical equations describing dissipation in the different regimes.

 \subsection{Super-resonant regime (red area in Fig.~\ref{Fig-regimes})} 
We start analysis of different regimes shown in Fig. \ref{Fig-regimes} 
with discussing super-resonant regime, 
\be   \gamma \ll \omega_2 \ll \omega_1 ,
\label{eq:super-cond}
\ee 
which can be realized in high-quality structures  or in structures with high electron concentration.

\begin{figure}[h!]
\centering
\includegraphics[width=8.6 cm]{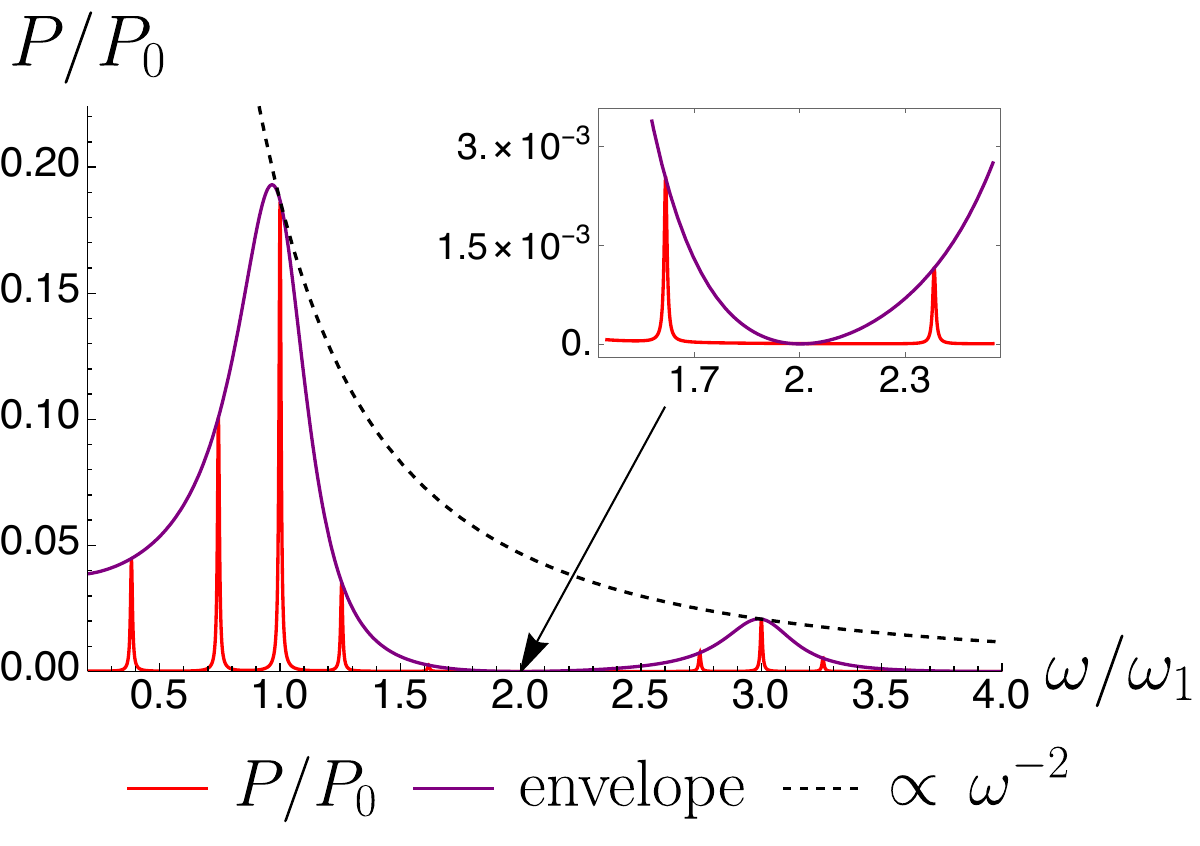}
\caption{ Frequency dependence of dissipation in the  super-resonance regime (red peaks) calculated by using Eq.~\eqref{Eq-maindis} for $s_2=0.2 s_1, ~\gamma=0.01 \omega_1,~ L_1 = L_2,~ \omega_c=0$.   Violet envelope is described by   Eq.~\eqref{eq-super-env}. Dashed line is proportional to $1/\omega^2.$   }
\label{Fig-envelope}
\end{figure}

Typical frequency dependence of dissipation in this regime calculated by using  Eq.~\eqref{Eq-maindis} is shown in Fig.~\ref{Fig-envelope}. One can see narrow resonances with the smooth envelope.   

Next, we describe Fig.~\ref{Fig-envelope} analytically.  
We will search for resonance harmonics of $\omega_1$ and $\omega_2.$ Therefore, having in mind  Eq~\eqref{eq:super-cond}, we  assume that $\omega \gg \gamma.$ We also expand $\Sigma$ over $\gamma,$
\be
\Sigma=\Sigma_0- i\frac{\gamma }{4} \mathcal L , 
\label{eq:sig0-iL}
\ee
where
\begin{align}
&\Sigma_0\!=\!s_1 \cot{\frac{\omega L_1}{2 s_1}}\!+\!s_2 \cot{\frac{\omega L_2}{2 s_2}}\!= \!\frac{Q_{\rm bright} (\omega)}{ \sin \left(\!\frac{\omega L_1}{2 s_1} \!\right) \sin \left(\!\frac{\omega L_2}{2 s_2} \!\right)}\!,
\label{Eq-sigma_bright}
\\
&\mathcal L = \mathcal L(\omega) = \frac{L_1}{ \sin^{2}{\left(\frac{\omega L_1}{2 s_1}\right)}} +  \frac{L_2}{ \sin^{2}{\left(\frac{\omega L_2}{2 s_2}\right)}}.
    \label{Eq-Lkrasivoe}
\end{align}
Then, one can simplify 
 Eq.~\eqref{Eq-maindis} as follows:
\be
\begin{aligned}
&P = \frac{P_0}{s_1^2 (L_1+L_2)} \frac{\gamma^2}{\omega^2
} 
\\
& \times \left[ (L_1 s_1^2 + L_2 s_2^2)   +  \frac{(s_1^2 - s_2^2)^2 \mathcal L  }{\Sigma_0^2 + \gamma^2 \mathcal L^2/16 }  \right].
\end{aligned}
\label{Eq-resdis}
\ee
Here, we introduced normalization constant
\begin{equation}
    P_0 = \frac{F_0^2 C s_1^2}{2 \gamma e^2} =  \frac{F_0^2 N_1}{2 m \gamma},
    \label{Eq-P0}
\end{equation}
which is the  quasistatic dissipation (for $\omega=0$) in the homogeneous 2D liquid ($s_2=s_1$, i.e. $N_1=N_2$) in the absence of the magnetic field.  Eq.~\eqref{Eq-resdis} does not apply at very small frequency, $\omega  \lesssim \gamma .$   so that the Drude peak   will be discussed below separately at Sec.~\ref{Sec-low_freq}.
As seen from  Eqs.\eqref{Eq-sigma_bright} and \eqref{Eq-resdis},  $P$  has resonances when $Q_{\rm bright}(\omega)=0,$ while solutions of equation $Q_{\rm dark} (\omega)=0$ do not show up in the excitation spectra. Hence, notations {\it bright} and {\it dark}.  Below in Sec.~\ref{Sec-light_modulation} we demonstrate that dark modes can show up due to the  optical modulation,  $h \neq 0$ and/or $K \neq 0.$   

Plasmonic resonant  frequencies $\omega_m$  are found from the condition $\Sigma_0(\omega_m)=0.$ 
For $\omega$ close to resonance frequency, $|\omega-\omega_m| \sim \gamma, $ 
one can neglect first term in the square brackets in  Eq.~\eqref{Eq-resdis}].  Within this resonance approximation we  get:
\begin{equation}
P(\omega \approx \omega_m) \approx \frac{A}{\omega_m^2} \frac{\gamma^2 \mathcal L} { \Sigma_0^2+ \gamma^2 \mathcal L^2/16},
\label{eq-Pestimate}
\end{equation}
where
\begin{equation}
    A = \frac{P_0 (s_1^2 - s_2^2)^2}{2   (L_1 + L_2) s_1^2 }.
    \label{Eq-A}
\end{equation}
Exactly at the $m-$th resonance, when $\omega=\omega_m,$ dissipation reads
\begin{equation} 
P_m  = \frac{16 A} {\omega_m^2 \mathcal L_m},
\label{Pm}
\end{equation}
where $\mathcal L_m= \mathcal L(\omega_m). $
This equation allows to find the envelope of the sharp resonances (see violet curve in Fig.~\ref{Fig-envelope}). 
To this end,  we  use 
condition of  the resonance [$Q_{\rm bright}(\omega) =0,$ or, equivalently, $\Sigma_0(\omega) =0$],   
\begin{equation}
 s_1 \cot{\frac{\omega L_1}{2 s_1}}+s_2 \cot{\frac{\omega L_2}{2 s_2}} = 0,
 \label{res1}
 \end{equation}
to express  $\cot({\omega L_2}/{2 s_2})$   
and substitute it into Eq.~\eqref{Eq-Lkrasivoe} in order to  finds values of $\mathcal L_m.$ Finally, substituting thus found $\mathcal L_m$ into Eq.~\eqref{Pm} and replacing $\omega_m \to \omega,$ we arrive at the following expression for the  envelope  
\be
P_{\rm env} \!=\!  \frac{16 A }{\omega^2}
\frac{\sin^2 \frac{\omega L_1}{2 s_1} }
{
L_1 \!+\! L_2 
\left[ \sin^2 \frac{\omega L_1}{2 s_1} \!+\!\left(\frac{s_1}{s_2}\right)^2 \cos^2 \frac{\omega L_1}{2 s_1}  \right] }.
\label{eq-super-env}
\ee

For $s_1 \gg s_2, $ this envelope  also contains resonances
at frequencies $\omega_n=(2n+1)\omega_1.$ Introducing $\delta \omega= \omega-\omega_n$ and assuming $|\delta \omega| \ll \omega_n, $
we find that   these  resonances obey
\be
P \approx \frac{8 P_0}{\pi^2}
\left(\frac{L_1}{L_1+L_2}\right)^2 \frac{1}{(2n+1)^2} \frac{1}{1+(\delta \omega /\Delta \omega)^2 },
\label{eq:Pdw}
\ee
where  
\be
\Delta w = \omega_2 \times \frac{2 \sqrt{L_2 (L_1 + L_2)}}{\pi L_1 },
\ee
and factor $1/(2n+1)^2$ arises  due to frequency dependence of $A \propto \omega^{-2}.$ 
Using Eq.~\eqref{res1} one can find positions of individual peaks (red peaks in  Fig.~\ref{Fig-envelope}) and maximal values, $P_{nm}$ in the fine structure described by integer $m$  within $n-$the peak of the envelope. For simplicity, we assume $L_1=L_2.$ Introducing notation $m =\delta \omega/\omega_2,$ we find from Eq.~\eqref{eq:Pdw}: 
\be
P_{nm}= P_0 \frac{2}{\pi^2}  \frac{1}{(2n+1)^2} \frac{1}{1+ m^2 \pi^2/8},
\ee
where values of  $m$  are not necessarily  integer and  are found from 
\begin{equation}
    \cot{\left[ \frac{\pi}{2} \left( N_n+ m \right) \right]} = \frac{\pi m}{2}.
\end{equation}
Here, $N_n = ({s_1}/{s_2}) \left(2 n + 1 \right) \gg 1.$ We notice that for half-integer values of $N_n,$  there is a solution with $m=0,$ i.e. exactly in the center of $n-$th peak of the envelope.  

\subsection{Resonant regime (blue area in Fig.~\ref{Fig-regimes}) }
As $\gamma$  increases for fixed $s_2/s_1,$ we enter the resonant regime:
\be     \omega_2 \ll \gamma  \ll \omega_1 
\label{eq:res-cond}
\ee 
 (here, we assume   strong coupling, $s_2/s_1 \ll 1$).  In this case,  red peaks shown in Fig.~\ref{Fig-envelope} overlap and the fine structure of the dissipation  disappears. On the technical level, the term     $\cot(\omega L_1/2 s_1) $ can be expanded over $\gamma$ just as in the super-resonant regime, while $\cot(\omega L_2/2 s_2) \to -i,$ so that Eq.~\eqref{sigma}  can be still written in the form  \eqref{eq:sig0-iL} with
$\Sigma_0 \approx s_1 \cot\left ( {\omega L_1}/{2 s_1}\right)$ and
$\mathcal L \approx L_1/ \sin^2\left({\omega L_1}/{2 s_1} \right) + 4 s_2/\gamma.$
The  term $4 s_2/\gamma$ entering expression for $\mathcal L$   is responsible for decay of plasmonic oscillations in the active regions due  to the coupling with   the overdamped plasmons in the   passive regions  \cite{Boubanga-Tombet2020}. Having in mind condition Eq.~\eqref{eq:res-cond}, one can neglect this term provided that $L_1$  is on the order of $L_2.$  
Within the resonance approximation, we get    
\begin{equation}
    P^{\rm res} = P_0 \sum_{n =0}^\infty \frac{\gamma^2 B_n}{(\omega- \omega_n^{\rm strong})^2+\gamma^2/4},
    \label{Eq:Pstrong}
\end{equation}
where 
\begin{equation}
    \omega_n^{\rm strong}=  (2 n + 1) \omega_1,  \label{Eq-w_strong} \end{equation} 
and
\begin{equation}
    B_n = \frac{2  L_1}{(L_1 + L_2) (1 + 2 n)^2 \pi^2}.
    \label{Eq-Bm}
\end{equation}
Physically, Eq.~\eqref{Eq:Pstrong} corresponds to  the plasmonic resonances in the independent active stripes.  The factor $L_1/(L_1+L_2)$  appears because we  calculate the  dissipation per unit length averaging over the whole sell of  the PC.  Up to this factor,  Eq.~\eqref{Eq:Pstrong} coincides with Eq.~(S29) from  Ref.~\cite{Boubanga-Tombet2020}.

We  notice that  only  modes with odd numbers $2 n+1$ are excited, while even modes $ 2 n $ are {\it dark} in the absence of the  optical modulation.

\subsection{Non-resonant regime (grey area in Fig.~\ref{Fig-regimes})}
\label{Sec-overdamped}
Non-resonant regime corresponds to the condition $\gamma \gg \omega.$ 
Then, $\Omega\approx \Gamma \approx \sqrt{\omega \gamma /2}.$ Assuming also that $\omega$ is not too small, $\omega \gg \omega_{1,2}^2/\gamma,$ we find $\Sigma \approx -i(s_1+s_2)$ (the case of very small $\omega$ will be discussed below in Sec.~\ref{Sec-low_freq}). Then, from Eq.~\eqref{Eq-maindis} we obtain
\begin{equation}
\begin{aligned}
&P_{\rm non{\text -}res}  \approx  \frac{C F_0^2 (L_1 s_1^2+L_2 s_2^2)}{2 e^2 \gamma (L_1+L_2)}
\\
&\times \left[1-\frac{\sqrt{2} (s_1-s_2)^2 (s_1 +s_2)}{(L_1 s_1^2 + L_2 s_2^2) \sqrt{\omega \gamma}} \right].
\end{aligned}
\label{Eq-nonresdis}
\end{equation}
In order to compare this equation with previously obtained  results for nonresonant regime  in single FETs \cite{Dyakonov1996,Veksler2006}  we take  the limit $s_2 \to 0 ,$ corresponding to independent active stripes separated by dielectric passive regions.  Then, Eq.~\eqref{Eq-nonresdis}  simplifies:
\begin{equation}
    P_{\rm non{\text -}res}(s_2=0) =P_0 \frac{L_1}{L_1+L_2}  \left( 1 - \frac{\sqrt{2} L_{1}^{*}}{L_1 } \right),
\end{equation}
where  $L_{1}^{*} = s_1/\sqrt{\omega \gamma} \ll L_1$  is the decay length of charge density oscillations in the ohmic regime, when plasma oscillations are overdamped   \cite{Dyakonov1996,Veksler2006}. Small correction to dissipation $\propto L_1^*/L_1$ comes from  narrow  layers with the width  $L_1^*$ near the boundaries of the active regions.

\subsection{Low frequencies: Drude peak}
\label{Sec-low_freq}
The equations  derived above are not applicable in the low frequency limit.
For arbitrary $\omega_{1,2}$,  and $\gamma,$ at very small $\omega$  ($\omega \ll \omega_2,$ $\omega \ll \gamma$),   Eq.~\eqref{sigma} yields
\be
\Sigma\approx \frac{2\left(\frac{s_1^2}{L_1}+\frac{s_2^2}{L_2}\right)}{\Omega + i \Gamma}.
\ee
Substituting this equation into Eq.~\eqref{Eq-maindis}, we obtain Drude peak 
\begin{equation}
    P_{\rm Drude} (\omega) = P_0 \frac{\gamma^2}{\gamma^2+ \omega^2} \xi 
    \label{PC-Drude_0}
\end{equation}
with
\be
\xi = \frac{1+L_2/L_1}{1+ L_2 s_1^2/L_1 s_2^2}. 
   \label{Eq-xi}
\ee
It is worth noting,  that this peak shows up in the dissipation in the super-resonant case, when $\omega_2 \gg \gamma.$  In the opposite resonant case, $\omega_2 \ll \gamma,$ the Drude peak overlap with resonances at frequency $ \omega_2$ and its harmonics.  Therefore, the peak is absent, while the value $P_{\rm Drude} (0)=P_0 \xi$  gives the value of dissipation at zero frequency for $\omega_2 \ll \gamma \ll \omega_1$ and $\omega \to 0.$  As follows from Eqs.~\eqref{PC-Drude_0} and \eqref{Eq-xi},   $P_{\rm Drude} \propto s_2^2$ for small $s_2.$   Physically, this happens because  passive regions become insulating for  $s_2 = 0$ thus blocking dc current and consequently the dissipation in the stationary case $\omega=0.$ 

Interestingly, if we  first put $s_2=0$ and  next consider limit of small $\omega,$ we get result which complements Eq.~\eqref{PC-Drude_0} for $s_2=0$:  
\begin{equation}
    P^{\rm strong} (\omega \to 0) = \frac{\pi^4}{120} \,  P_0 \frac{\gamma^2 \omega^2}{\omega_1^4} \frac{L_1}{L_1+L_2}.
\end{equation}

\subsection{Weak coupling,  $s_1 \approx s_2$}
\label{Sec-weak_coupling}
In the weak coupling case,  resonant regime is absent, so that with increasing $\gamma$ there is a transition from the super-resonant regime to non-resonant one (see Fig.~\ref{Fig-regimes}).  

For  $\gamma<\omega_{1,2},$ the system is in the super-resonant regime. In this case, we put $s_2 \to s_1$ everywhere except the factor $(s_1-s_2)^2$ in the   second term in the square bracket of Eq.~\eqref{Eq-maindis}.  That is the term that yields resonances at $\omega_n^{\rm weak}$ given by Eq.~\eqref{Eq-w_weak}.  Using resonance approximation for each peak, after some algebra we write response as follows      
 
\begin{equation}
    P^{\rm weak} = P_{\rm Drude}+P_0 \sum_{n = 1}^\infty \frac{\gamma^2 A_n}{(\omega- \omega_n^{\rm weak})^2+\gamma^2/4},
    \label{Eq-P_weak}    
\end{equation}
where $P_{\rm{Drude}} \approx  P_0 \gamma^2/(\omega^2+\gamma^2)$ is the Drude dissipation for  $s_1=s_2$ 
and 
 \begin{equation}
    A_n =\frac{4 (s_1-s_2)^2 \left( 1- (-1)^n \cos{\left[\frac{ n  k (L_1-L_2)}{2 } \right]} \right )}{(\omega_n^{\rm weak})^2 (L_1+L_2)^2}
    \label{Eq-An}
\end{equation}
is a dimensionless amplitude  of $n-$th harmonic. For  $L_1=L_2=L/2,$ the  amplitudes of even harmonics  vanish:  $A_{n = 2m} = 0$. The  amplitude $A_n$  decays with $n$ as $ 1/(\omega_n^{\rm weak})^2 \propto 1/n^2$ 
just as in the strong coupling case. 

With increasing  $\gamma$ above $ \omega_1 \approx \omega_2$, we arrive at 
overdamped  non-resonant regime, where only   broad Drude resonance is visible in dissipation.

 We should also  make a comment regarding  dark and bright modes in the weak coupling regime.    Equations ~\eqref{Eq-P_weak} and \eqref{Eq-An} were found in the second perturbation order 
with respect to $\delta s=s_1-s_2.$  For $\delta s=0,$ $Q_{\rm bright} (\omega) \equiv Q_{\rm dark} (\omega),$ so that  frequencies of dark and bright modes coincide and are given by Eq.~\eqref{Eq-w_weak}. Let us consider $n=1.$  Then, $$\omega_1^{\rm bright} =\omega_1^{\rm dark} = s_1 k, \qquad \text{for}~s_1=s_2.  $$ 
Using Eqs.~\eqref{Eq-Qbright} and \eqref{Eq-brackets}  one can  find  that dark and bright modes split when $\delta s \neq 0.$  In particular, for $L_1=L_2=L/2$, splitting $\delta \omega=\omega_1^{\rm bright} -\omega_1^{\rm dark}$ reads
\be
\delta \omega \approx\frac{\pi \delta s^2}{s_1 L}.
\label{Eq-delta_w2}
\ee
For $L_1 \neq L_2$ difference appears at first order of $\delta s$:
\be
\delta \omega \approx \frac{\sqrt{2} \delta s}{L_1+L_2} \sqrt{ \sin^2{\frac{2 \pi L_1}{L_1+L_2}}+ \sin^2{\frac{2 \pi L_2}{L_1+L_2}}}.
\label{Eq-delta_w}
\ee

\begin{figure}[h!]
\centering
\includegraphics[width=8.25 cm]{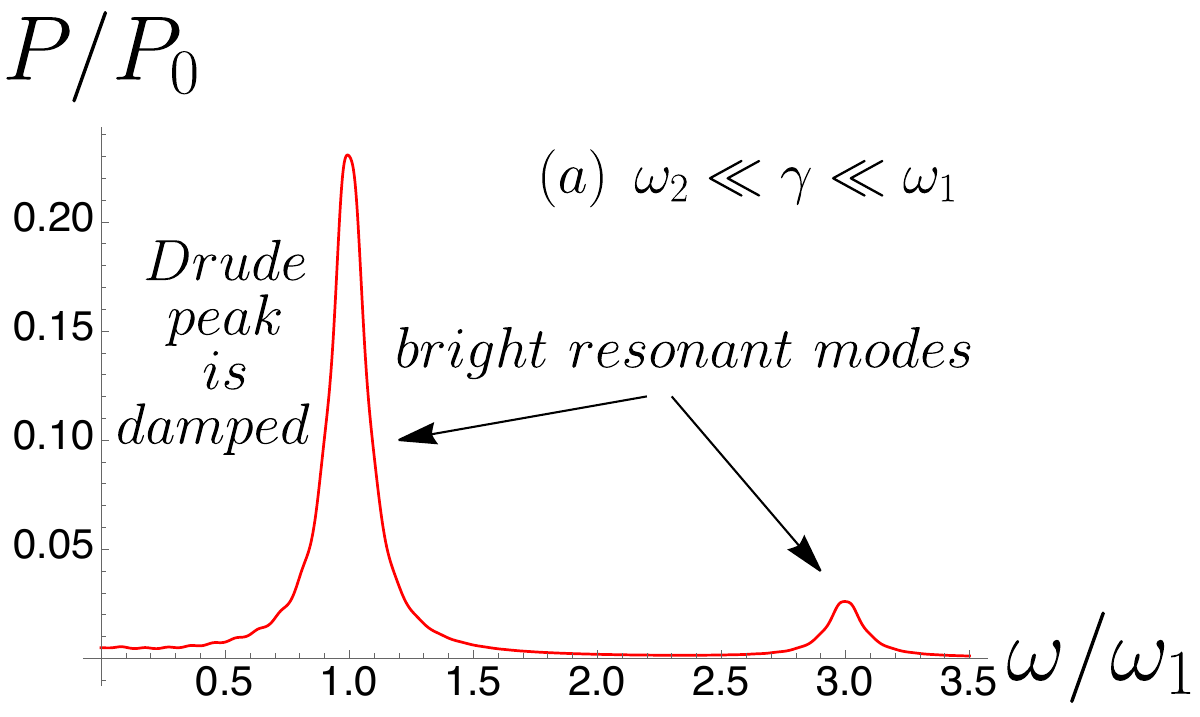}
\includegraphics[width=8.25 cm]{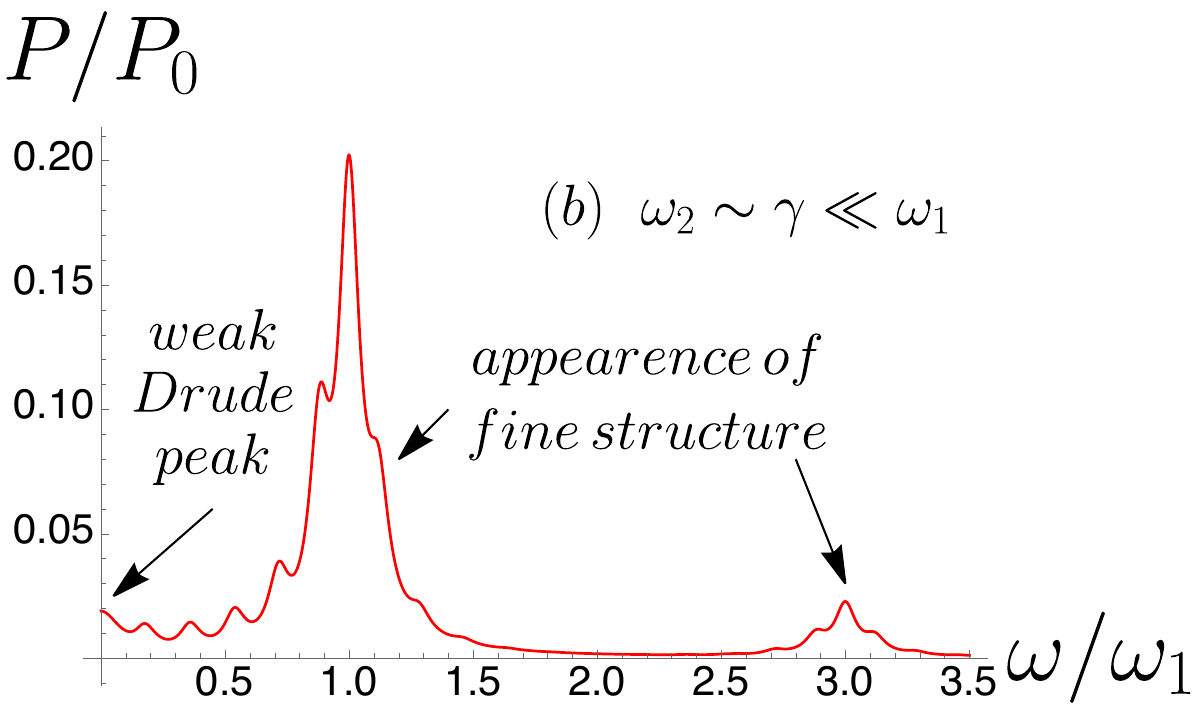}
\includegraphics[width=8.25 cm]{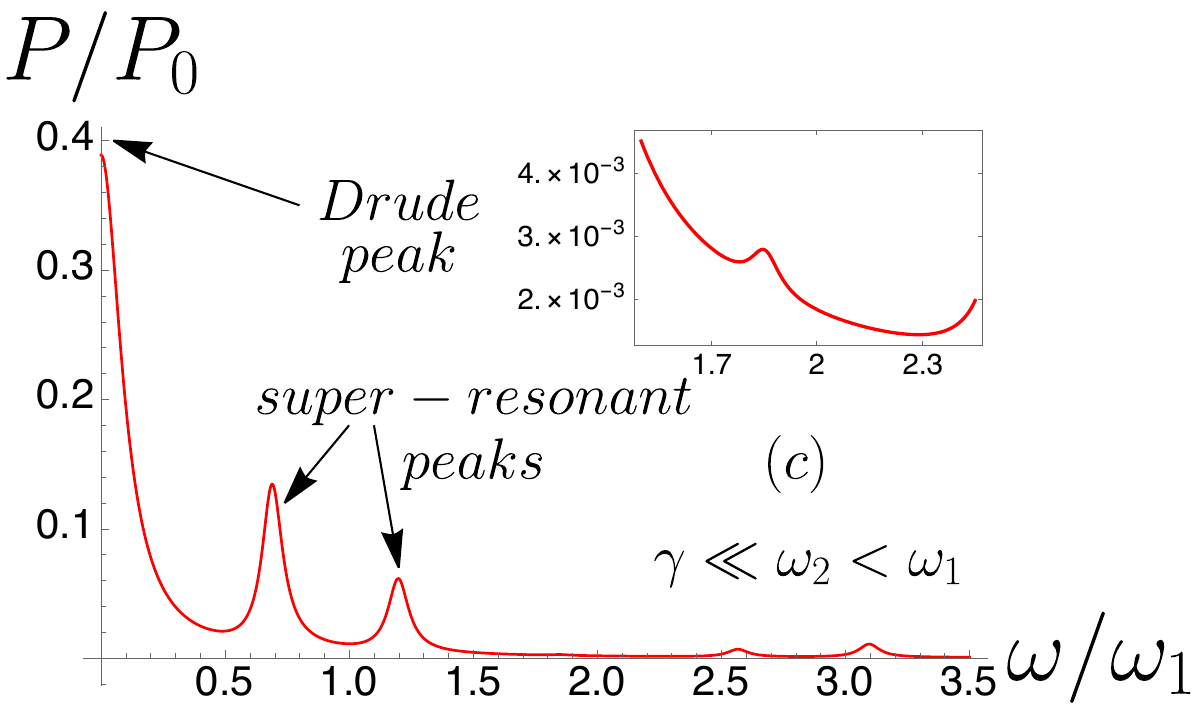}
\includegraphics[width=8.25 cm]{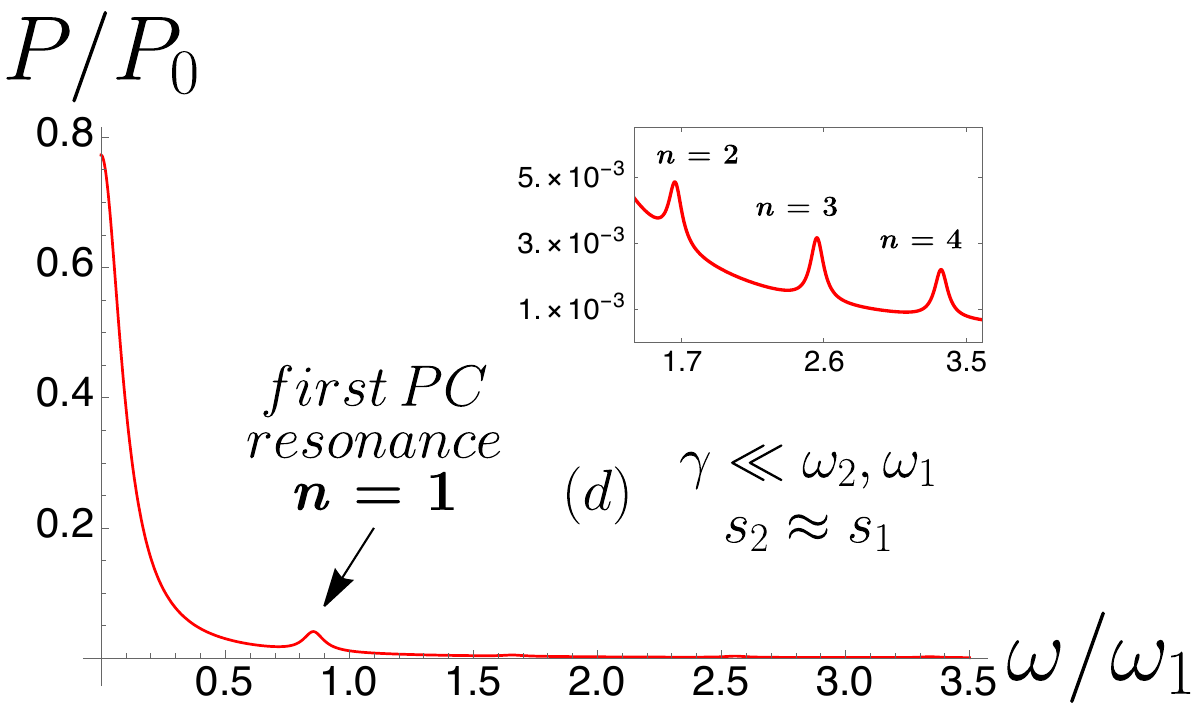}
\caption{ Evolution of dissipation ${P}/{P_0}$ with increasing  $s_2/s_1$ for fixed $\gamma=\omega_1/10$  and $L_2 = 1.1 L_1$ (horizontal cross section of Fig.~\ref{Fig-regimes}): (a) resonant regime, $s_2/s_1=1/20,$ (b) crossover between resonant and super-resonant regime,  $s_2/s_1=1/10,$ (c) developed super-resonant regime,   $s_2/s_1=1/2,$ (d) weak coupling regime, $s_2/s_1 = 4/5.$ Note difference in range of $P/P_0$.}
\label{Fig-evolution}
\end{figure}

Such small corrections to the resonant frequencies   were neglected in Eq.~\eqref{Eq-P_weak}. In order to probe experimentally the splitting $\delta \omega$  in the weak-coupling regime,  one should use very clean structures where $\delta \omega > \gamma.$

\subsection{Illustration of different regimes  }
\label{Sec-calculations}
Next, we present  several  plots, found using exact   equation \eqref{Eq-maindis} to analyse frequency dependence of the dissipation in different regimes.

In  Fig.~\ref{Fig-evolution}, in order to illustrate transition from resonant regime to super-resonant one,   we fix $\gamma$ (horizontal cross-section of Fig.~\ref{Fig-regimes}) at sufficiently small value, $\gamma \ll \omega_1,$ and  change $s_2$ (and consequently $\omega_2$) in a sufficiently wide range covering both regimes. 
We start with very low values of $\omega_2$ corresponding to resonant regime ($\omega_2\ll \gamma$)   and strong-coupling case ($\omega_2 \ll \omega_1 $)      (Fig.~\ref{Fig-evolution}a), and then increase $\omega_2.$  The boundary between resonant and   super-resonant regimes corresponds to the case $\omega_2 \sim \gamma$ (Fig.~\ref{Fig-evolution}b), while  developed super-resonant regime is shown in  Fig.~\ref{Fig-evolution}c. The latter  two panels, (b) and (c),   correspond to strong and intermediate coupling, respectively.  Finally, with further increase of $\omega_2$ we arrive at   weak coupling regime, which is illustrated in Fig.~\ref{Fig-evolution}d. 

Let us discuss different panels in   Fig.~\ref{Fig-evolution} in more detail.      In Fig. \ref{Fig-evolution}a 
dissipation shows  resonances $\omega^{\rm strong}_m = (2 m+1)\omega_1$ determined by  bright  plasmons in active region (dark active plasmons corresponds to even harmonics of $\omega_1$).   
 Physically, this regime  corresponds to excitation  of independent bright  plasmonic resonances in  active regions. ``Passive'' plasmonic resonances  strongly overlap and  do not show up because of condition $\gamma \gg \omega_2.$
 The Drude peak is also absent in this case, since  low conducting  passive region block the current in the stationary dc limit, $\omega \to 0.$     

Panel (b) shows intermediate regime -- between blue and pink regions at Fig. \ref{Fig-regimes}. The  resonances  at $\omega=\omega^{\rm strong}_m$, start to split, and fine structure corresponding to passive region appears. Specifically,  bright ``passive'' resonances at   $\omega=2 n \omega_2$  with small amplitude appears on the top  of  $\omega^{\rm strong}_m $ resonances.

Panel (c) illustrates two features of    the well developed super-resonant regime, $\omega_2 \gg \gamma $:  (i) ``active'' resonances  split into  ``passive'' ones and (ii) the amplitude of the  Drude peak  strongly increases.  

Finally,  panel (d) shows regime of weak coupling, when 2D electron liquid is  modulated weakly. Position of resonances well described with $\omega^{\rm weak}$, but the resonance   amplitudes are  small,  $\propto (s_1-s_2)^2,$  as compared to the dominating  Drude peak. As expected, this is the only peak that survives in the limit $s_2  \to  s_1$. 
  
\begin{figure}[h!]
\centering
\includegraphics[width=8.4 cm]{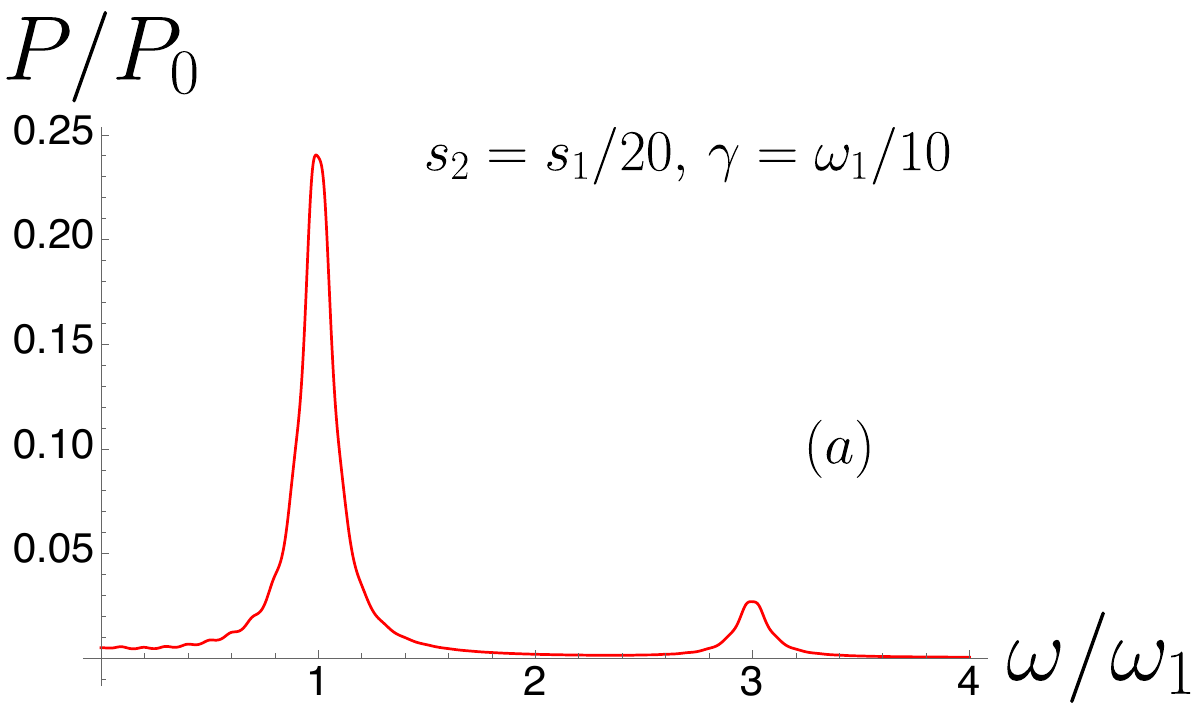}
\includegraphics[width=8.4 cm]{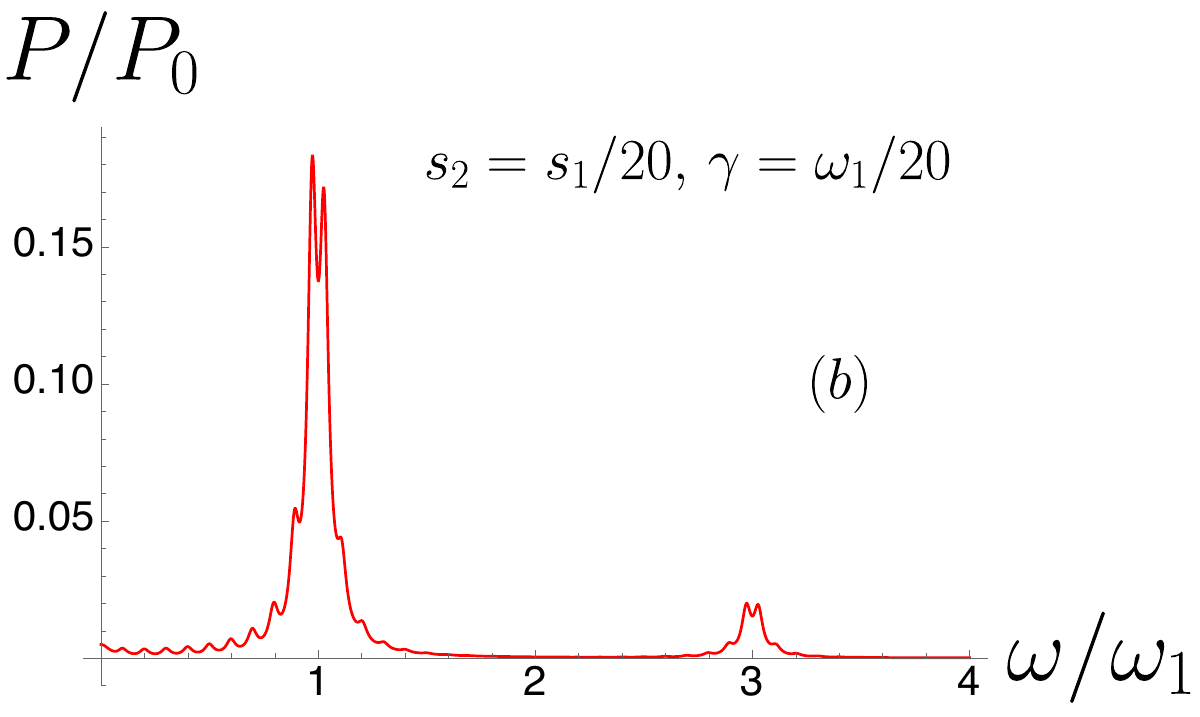}
\includegraphics[width=8.4 cm]{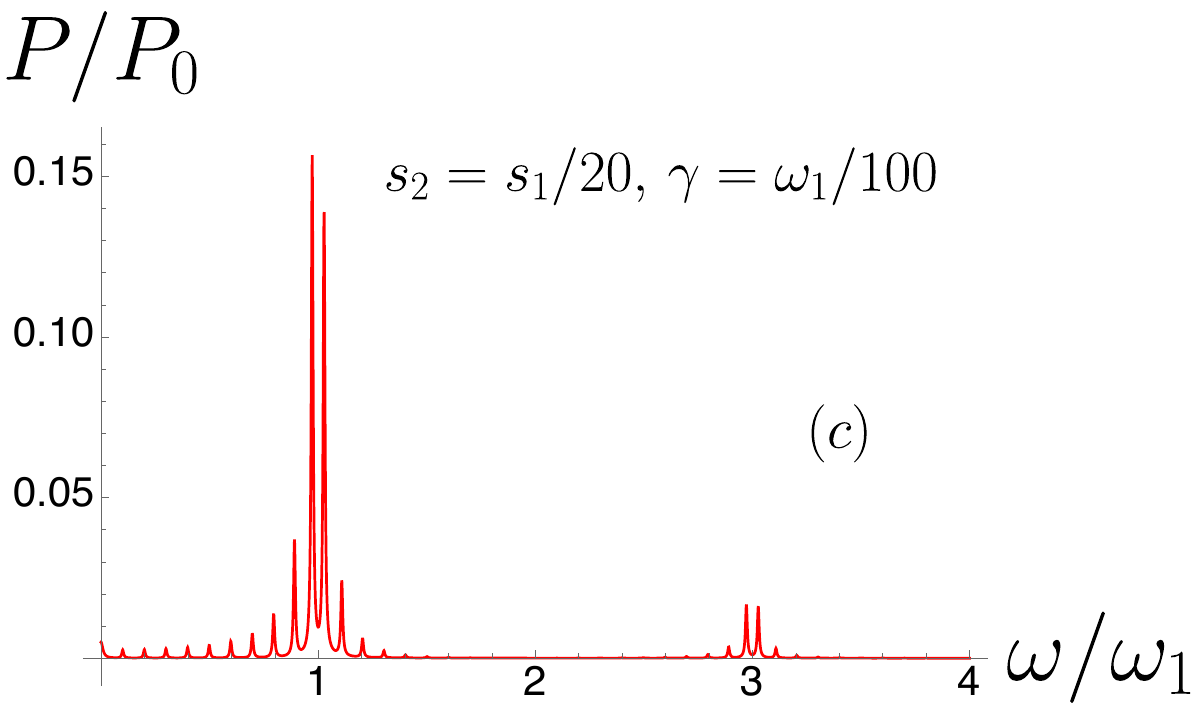}
\includegraphics[width=8.4 cm]{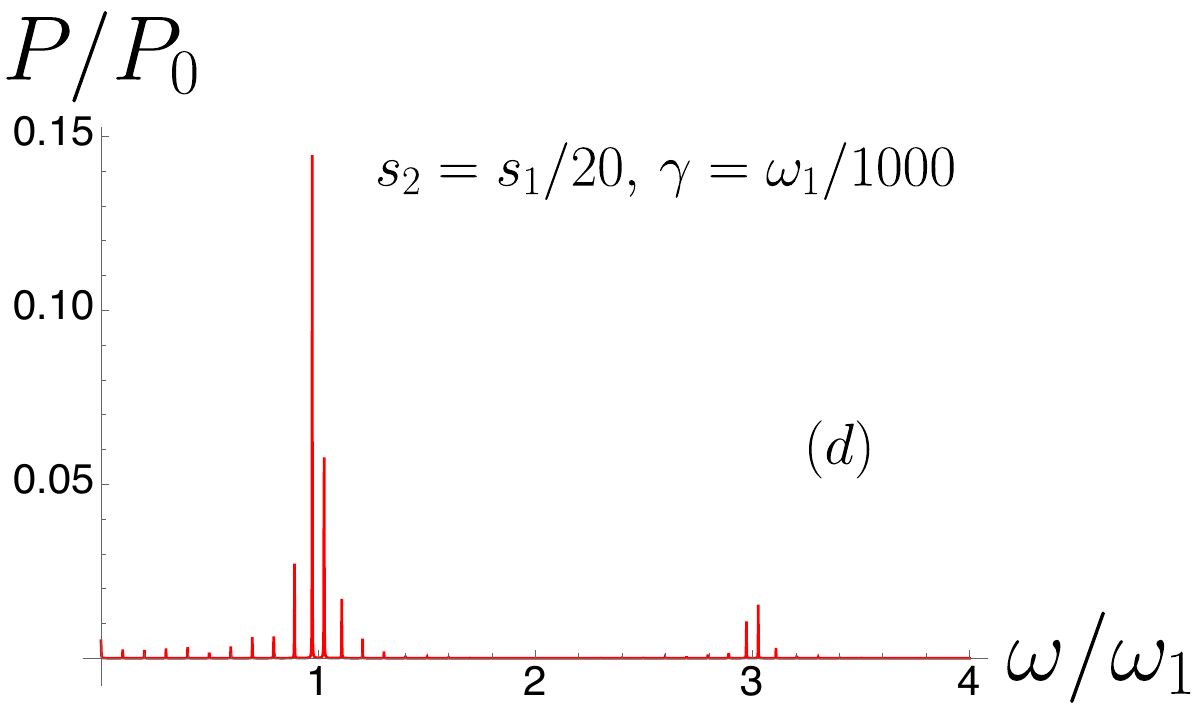}
\caption{ Evolution of frequency dependence of the dissipation  and appearance of fine super-resonant structure with decreasing $\gamma.$    Here,  $L_1 = L_2,~s_2/s_1=1/20,$ and the damping rate $\gamma$ decreases from panel (a) to panel (d)  (horizontal cross-section of Fig.~\ref{Fig-regimes}).  To avoid confusion, we note that $P_0 \propto 1/\gamma$ [see  Eq.~\eqref{Eq-P0}]. }
\label{Fig-evolution_gamma}
\end{figure}

Appearance of fine structure in ``active'' resonances is better seen  in Fig.~\ref{Fig-evolution_gamma}, where we  fix $s_2$ at the low value, $s_2/s_1=1/20,$  corresponding to the strong coupling case, and   study evolution of $P$ with decreasing of $\gamma.$ We again clearly see transition from the resonant to super-resonant  regime. Envelope of the fine structure show resonances at frequencies $\omega_m^{\rm strong} $.     

\section{Non-homogeneous excitation}
\label{Sec-light_modulation}
Above, we discussed homogeneous optical  excitation of inhomogeneous electron liquid.  Next, we take into account various types of  optical modulation.   
\subsection{Non-zero in-plane momentum  $K.$}
\label{Sec-PC_with_K}

\begin{figure}[h!]
\raisebox{0.01\height}{\includegraphics[height=2.6in]{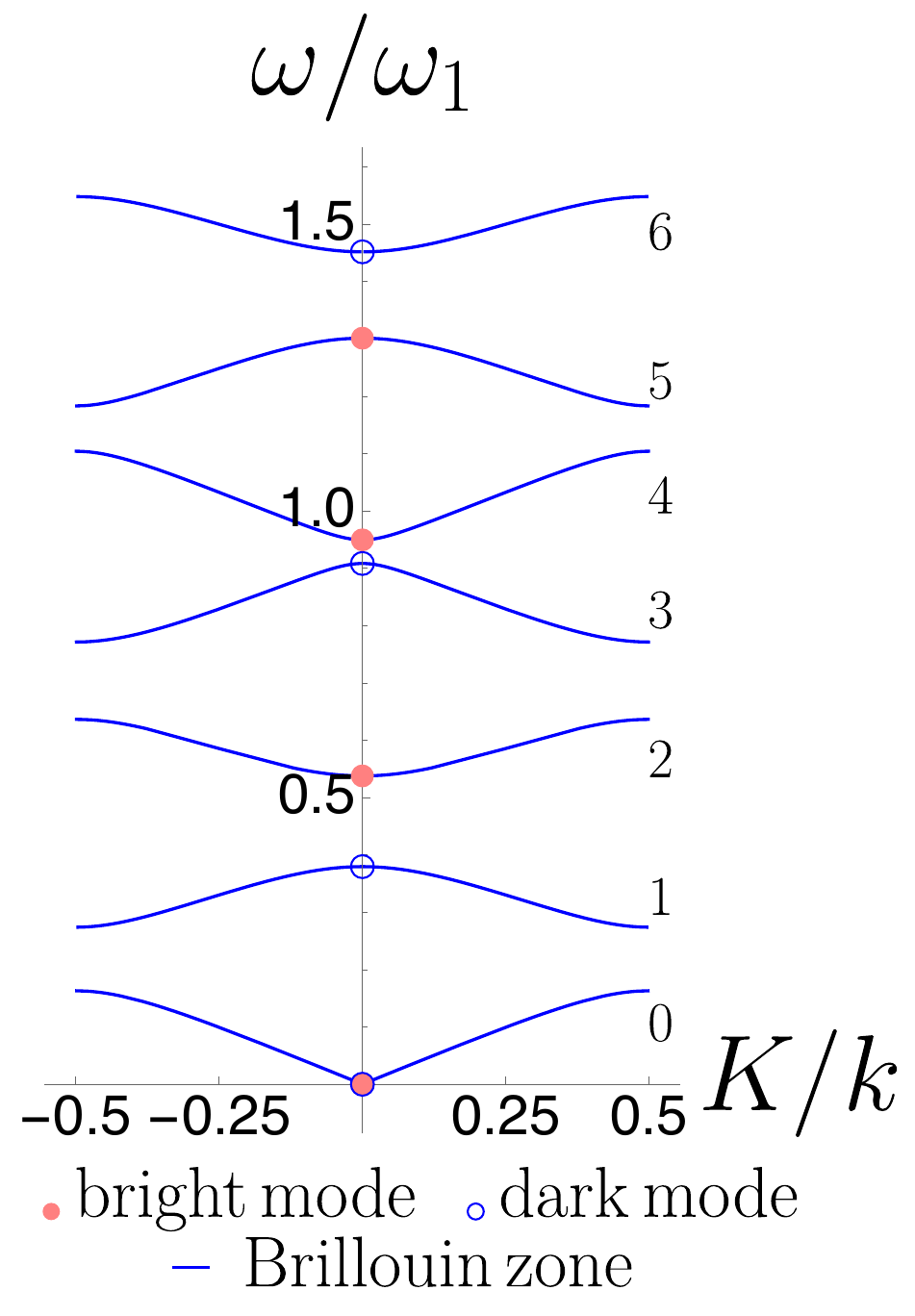}}
\hspace*{-0.3in}
\raisebox{-0\height}{\includegraphics[height=2.9in]{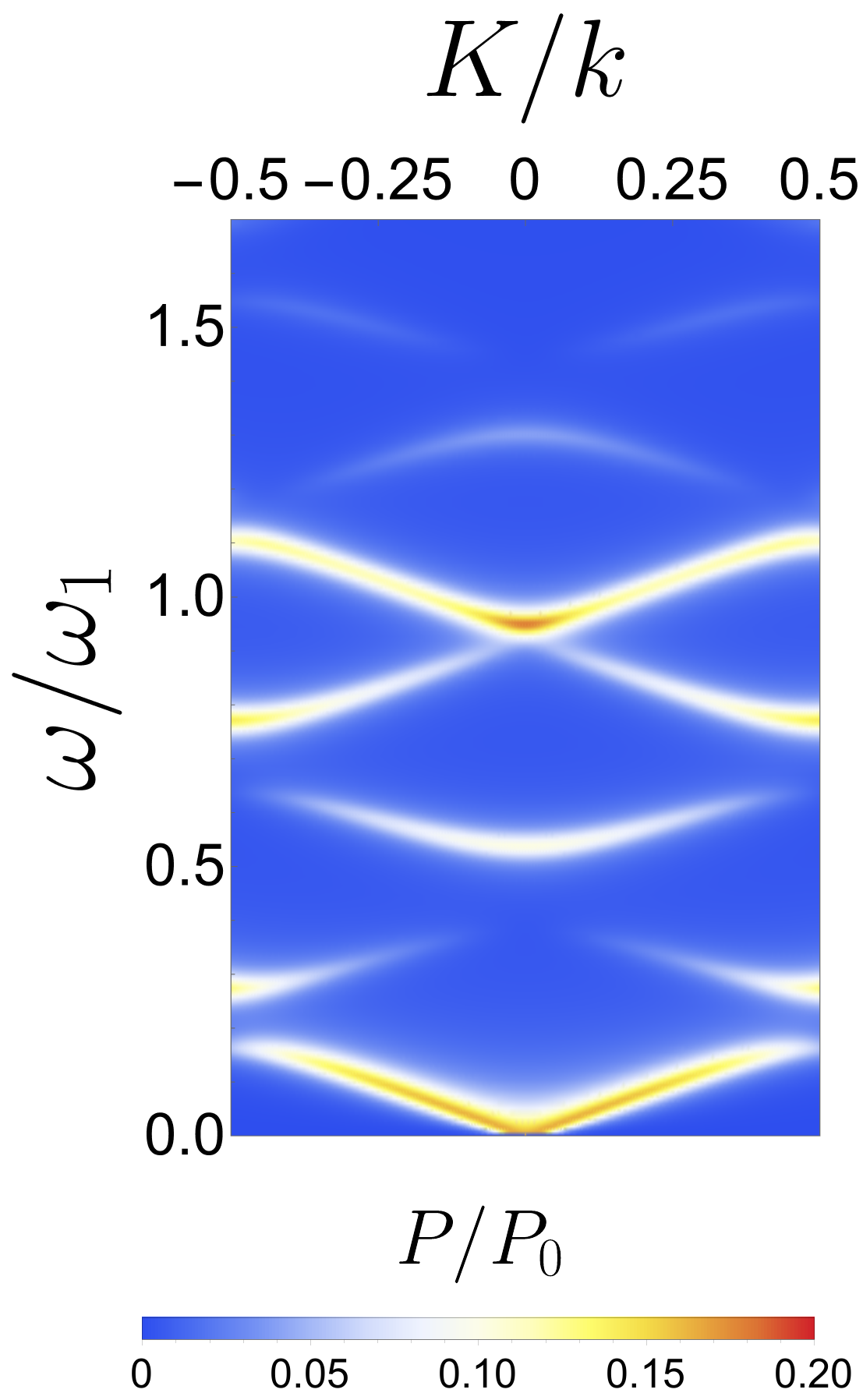}}
\caption{
Left panel:  spectrum of the ideal (i.e. with $\gamma=0$)  PC for $s_2 = 0.3~ s_1, ~L_1=L_2.$  Bright and dark modes are shown at  $K=0$  by thick red points  and open circles, respectively. Right panel: 
heatmap of the dissipation in the $(K,\omega)$ plane   for the PC with the same $s_{1,2},~ L_{1,2},$      and small momentum relaxation rate corresponding to the super-resonant regime, $\gamma =0.05 ~\omega_1$.   }
\label{Fig-darkstate2}
\end{figure}

 In this section, 
 we consider  modulation with non-zero in-plane momentum $K$  [see Eq.~\eqref{Eq-traveling_wave}].   Particularly, non-zero $K$ appears  when   incoming radiation   has  non-zero angle of    incidence.  Such an excitation can be  used to probe the dark states as was recently demonstrated  for visible light scattering on meta-surface of metallic particles   \cite{Hakala2017}.

We will show that     slow spatial modulation with  $K (L_1+L_2) \ll 1$ 
leads to the excitation of dark resonances with small amplitude,
$P(\omega_n^{\rm dark}) \propto K^2$.

\begin{figure}[h!]
\includegraphics[width=8.6 cm]{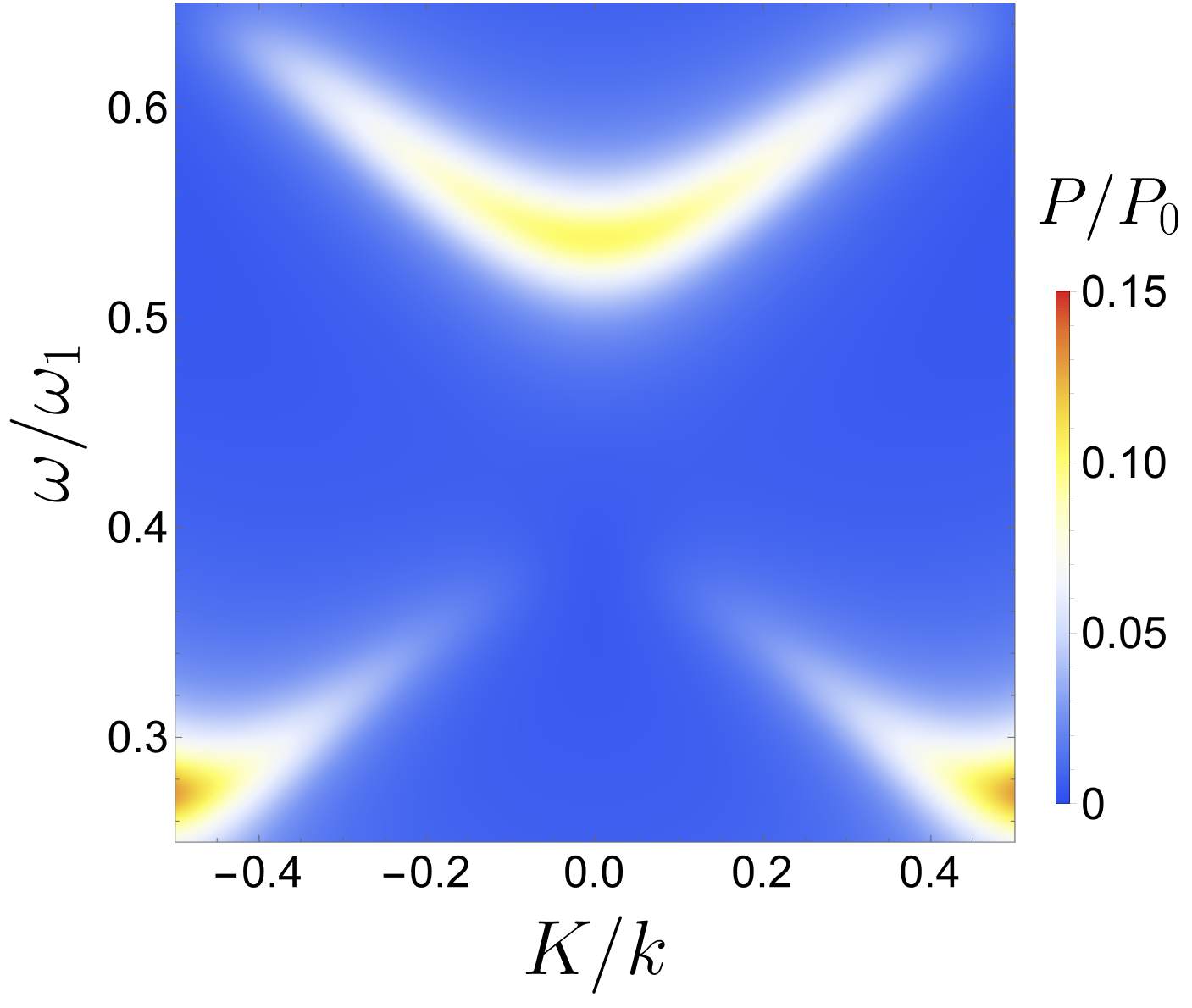}
\caption{ Illustration of different behavior of bright and  dark modes in the vicinity of point $K=0$ for bands  $1$ and $2$ (numeration corresponds to the left panel of Fig.~\ref{Fig-darkstate2}).  Heatmap  is shown in the interval  $K \in {(-k/2,k/2)},$ in the super-resonant regime  for the same parameters as in  the right panel of  Fig.~\ref{Fig-darkstate2} except  $\gamma=0.1~ \omega_1$. }
\label{Fig-darkstate}
\end{figure}

The  solution of Eqs.~\eqref{Eq-Navier_Stokes} and \eqref{Eq-continuity}  
with the electric field given by Eq.~\eqref{Eq-traveling_wave} is quite  
similar to the case of homogeneous field. Basic idea of calculation is  described in Appendix  
\ref{app:K-exp}.   However, the obtained analytical expression  for $P$ is too cumbersome to present it here.  
 This solution  allows us  to calculate a heat map  of dissipation in $(\omega, K)-$ space for arbitrary $s_2/s_1$ and $\gamma/\omega_1$  and discuss its features. Also,  we present below analytical expressions for simple limiting cases.

\begin{figure*}
\centering
\hspace*{-0.1in}
\includegraphics[width=0.315\textwidth]{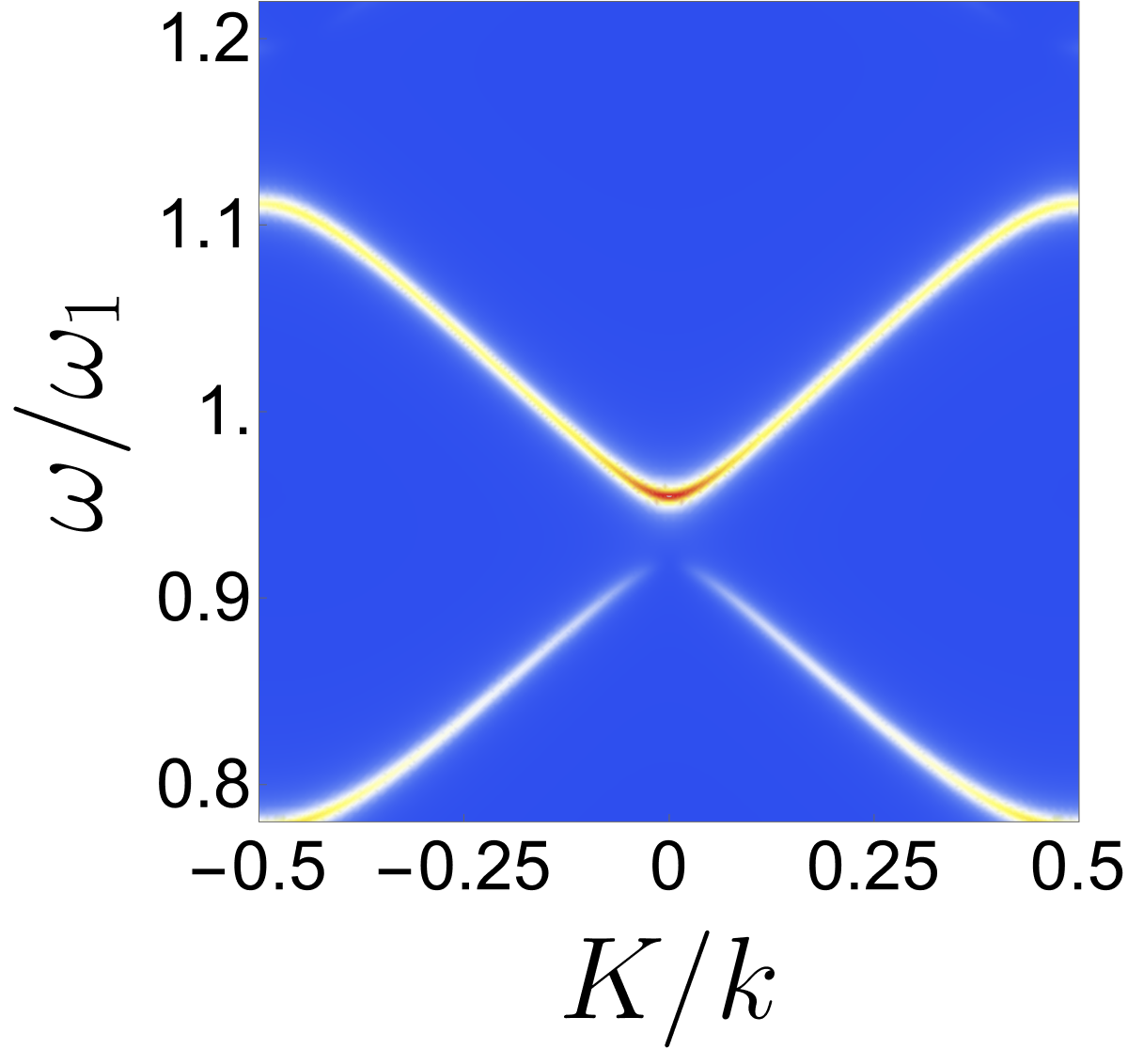}
\includegraphics[width=0.315\textwidth]{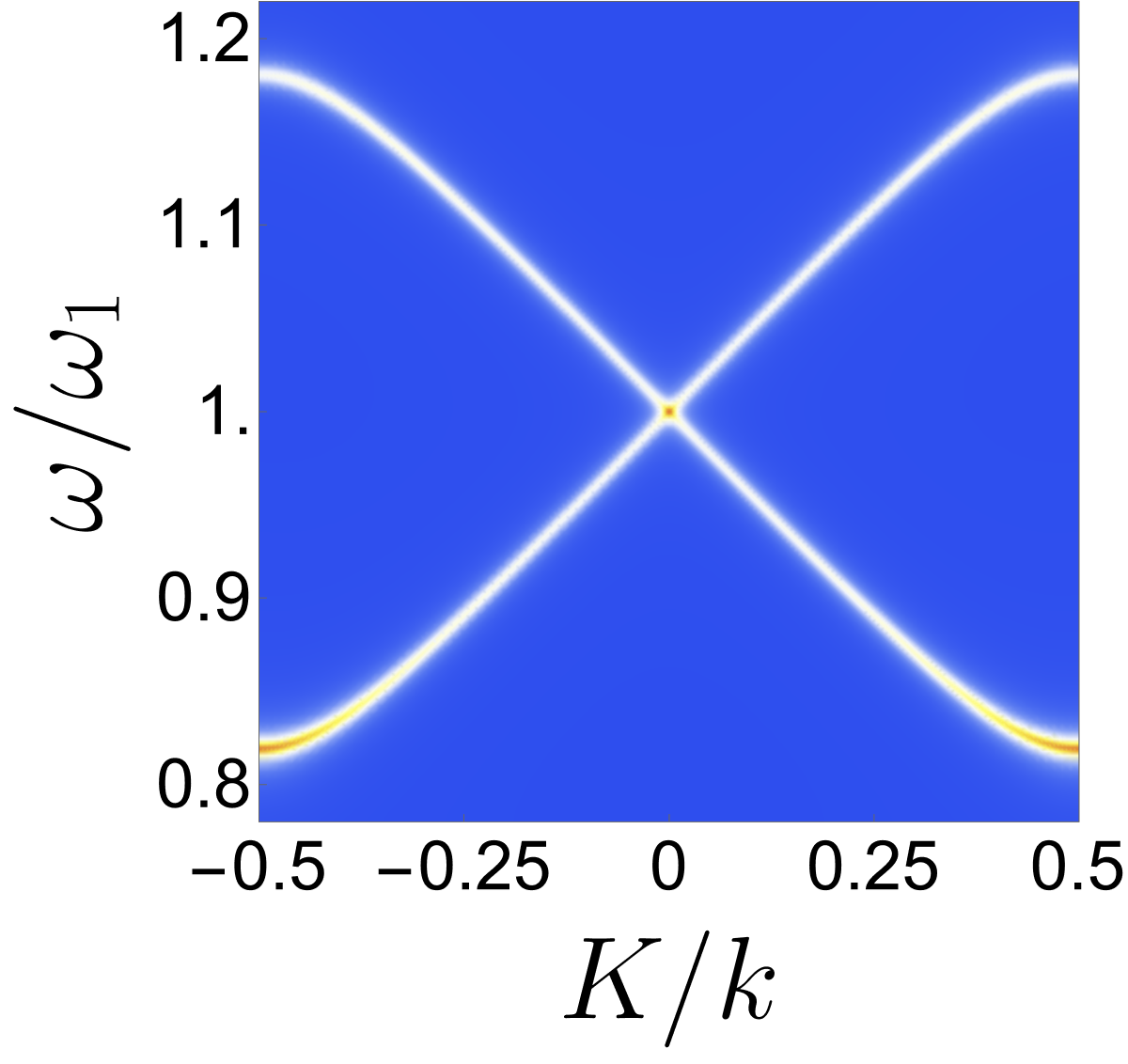}
\includegraphics[width=0.38\textwidth]{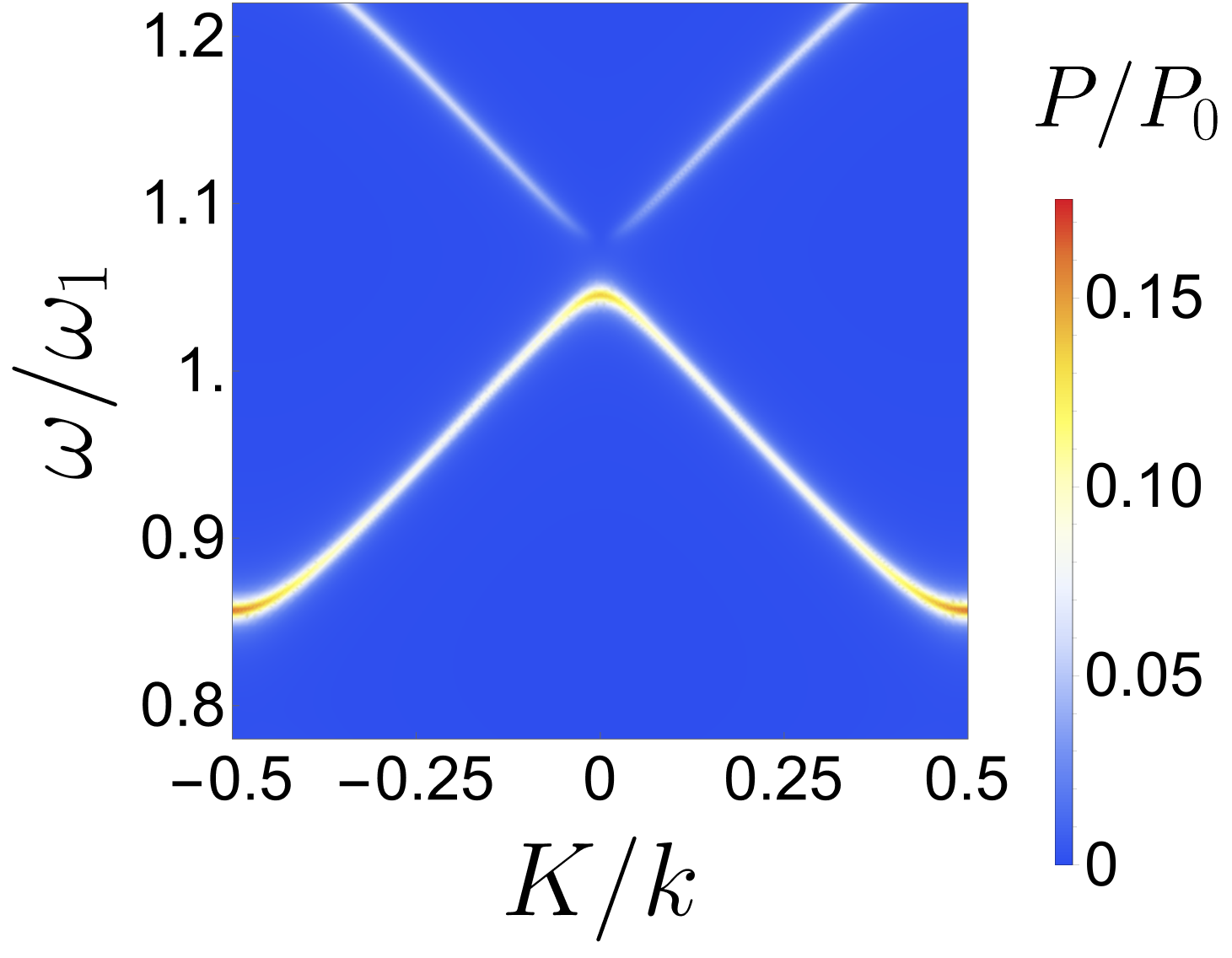}
\caption{ Interchange of the bright and dark states for $3$  and $4$ bands (numeration corresponds to the left panel of Fig.~\ref{Fig-darkstate2}).   Dissipation  map is plotted for  $L_{1}=L_{2}$  and  $s_2$ close to the value $s_1/3,$ corresponding to the intersection point with $n=0,m=1$ and $\omega=\omega_1$  according to Eq.~\eqref{Eq-cond1} [see intersection of red and grey lines in  Fig.~\ref{Fig-natfreq} plotted for $K=0$]. For better visualization of the intersection  the momentum relaxation rate is chosen smaller than  in    Fig.~\ref{Fig-darkstate2} and Fig.~\ref{Fig-darkstate},  $\gamma = 0.01 \omega_1.$ 
The ratio  $s_2/s_1$  changes from left to right as  $1/3 - 0.03,~1/3 ,$ and  $1/3+0.03.$  As seen, dark state ``jumps'' from  down band  to  the up one at $s_2/s_1=1/3.$  
}
\label{Fig-dark_state_jump}
\end{figure*}

In the  left panel of Fig.~\ref{Fig-darkstate2} one can see spectrum of the ideal (i.e. with $\gamma=0$)  PC for $s_2 = 0.3 s_1, ~L_1=L_2$ (left panel) and heatmap of the dissipation in the $(K,\omega)$ plane   for the PC with the same $s_{1,2},~ L_{1,2},$      and small momentum relaxation rate corresponding to super-resonant regime   (right panel).    Bright modes are shown in the  left panel of     Fig.~\ref{Fig-darkstate2}  by thick red points, while dark modes are shown by open circles.  
The heatmap in the right panel of     Fig.~\ref{Fig-darkstate2}  reproduce spectrum  with lines broadened due to momentum relaxation. Most important feature of this heat map is emerging  of ``dark'' resonances at $K \neq 0.$   For better illustration of this point we plotted in  the Fig.~\ref{Fig-darkstate} two pass bands   $1$ and $2$  (numeration according  to left panel of Fig.~\ref{Fig-darkstate2}).  One can see that  lower  band  is dark for $K=0$ and shows up in the dissipation only for $K \neq 0.$

Returning to the analysis of the left panel of  Fig.~\ref{Fig-darkstate2} we notice that spectrum  changes with changing $s_2/s_1,$ so that frequencies of bright and dark modes evolve and can intersect each other as was shown above in  Fig.~\ref{Fig-natfreq}. The possibility  of such intersections  is  illustrated  in Fig.~\ref{Fig-darkstate2}, where modes $3$ and  $4$ are very close to each other at $K=0.$  Exact intersection happens for $s_2= s_1/3$ [this intersection corresponds to $n=0, m=1$ and $\omega=\omega_1$ in  Eq.~\eqref{Eq-cond1}].  The evolution of the spectrum with $s_2$  in the vicinity of the value $ s_1/3$ is shown in  more detail in Fig.~\ref{Fig-dark_state_jump}, where bands $3$ and $4$ are shown for
$s_2/s_1$ slightly below   $1/3$, exactly  $1/3$ and  slightly above $1/3.$  As seen from this figure,   two bands approach,  then touch each other, and finally, diverge. Importantly, bright and dark points  interchange at $s_2/s_1 = 1/3.$   
Analyzing    Fig. \ref{Fig-darkstate2} we also notice     that there is no simple rule that prescribes   a spectrum branch to have or not to have dark state at $K = 0$. In the first  several bands $n = 0,1,2,3,4$    odd modes have dark states, while even modes have bright ones. However,  mode with $n=5$ also has bright state at $K=0.$

Appearance  of dark modes in the dissipation and, consequently, in  the transmission spectrum can be understood on the example of very strong coupling, $s_2 \to 0.$  This case   allows a simple analytical solution. As we discussed above, for $s_2=0$ oscillations in the different active stripes   are independent.   However, the external  field acting on electron liquid  in different stripes varies from stripe to stripe  and is inhomogeneous  within each stripe because of non-zero $K.$ 

Taking expressions for velocity and concentration within a stripe from  Appendix \ref{app:K-exp},  assuming ac current to be zero at the boundary between active and passive regions, and summing   over all active stripes we arrive at the following equation for  average dissipation per unit length 
\begin{align}
    P = \frac{L_1}{L_1+L_2} P_0 \times \sum_{n=1}^\infty  \frac{\gamma^2 E_n }{(\omega- n \omega_1)^2+\gamma^2/4},
\end{align}
where 
\begin{align}
E_n &= \frac{n^2 \pi^2  \left[1 - (-1)^n \cos{K L_1}\right]}{  ( n^2 \pi^2-K^2 L_1^2)^2} 
\nonumber
\\
&\approx  \frac{  \left[1 - (-1)^n \cos{K L_1}\right]}{  n^2 \pi^2} . 
\label{eq-En}
\end{align}
For $K\to 0$ only terms with odd $n$ survive in this sum and we restore Eq.~\eqref{Eq-Bm}, which yields the sum over  bright resonant modes.    For $K \ll 1,$ we find   $$E_{2 l+ 1} \approx  \frac{ 2-K^2 L_1^2/2}{\pi^2 (2 l+1)^2}, \, \, E_{2 l} \approx \frac{K^2 L_1^2}{2 \pi^2 (2 l)^2}.$$ 

Before closing this section we note that envelope  of the  resonant peaks decay with $n$ as $1/n^2.$  This agrees with the formula of dissipation in  homogeneous electron liquid,  $s_1=s_2,$
excited  by the wave with frequency $\omega$ spatially modulated with wave-vector $K$: 
\begin{equation}
    P = P_0 \frac{\gamma^2 \omega^2}{\gamma^2 \omega^2 + (\omega^2-K^2 s_1^2)^2}. 
\end{equation}
 For $\omega \gg (\gamma,s_1 K)$ this equation scales as  $ \propto 1/\omega^2.$ 
 
\begin{figure}[h!]
\centering
\includegraphics[width=8.1 cm]{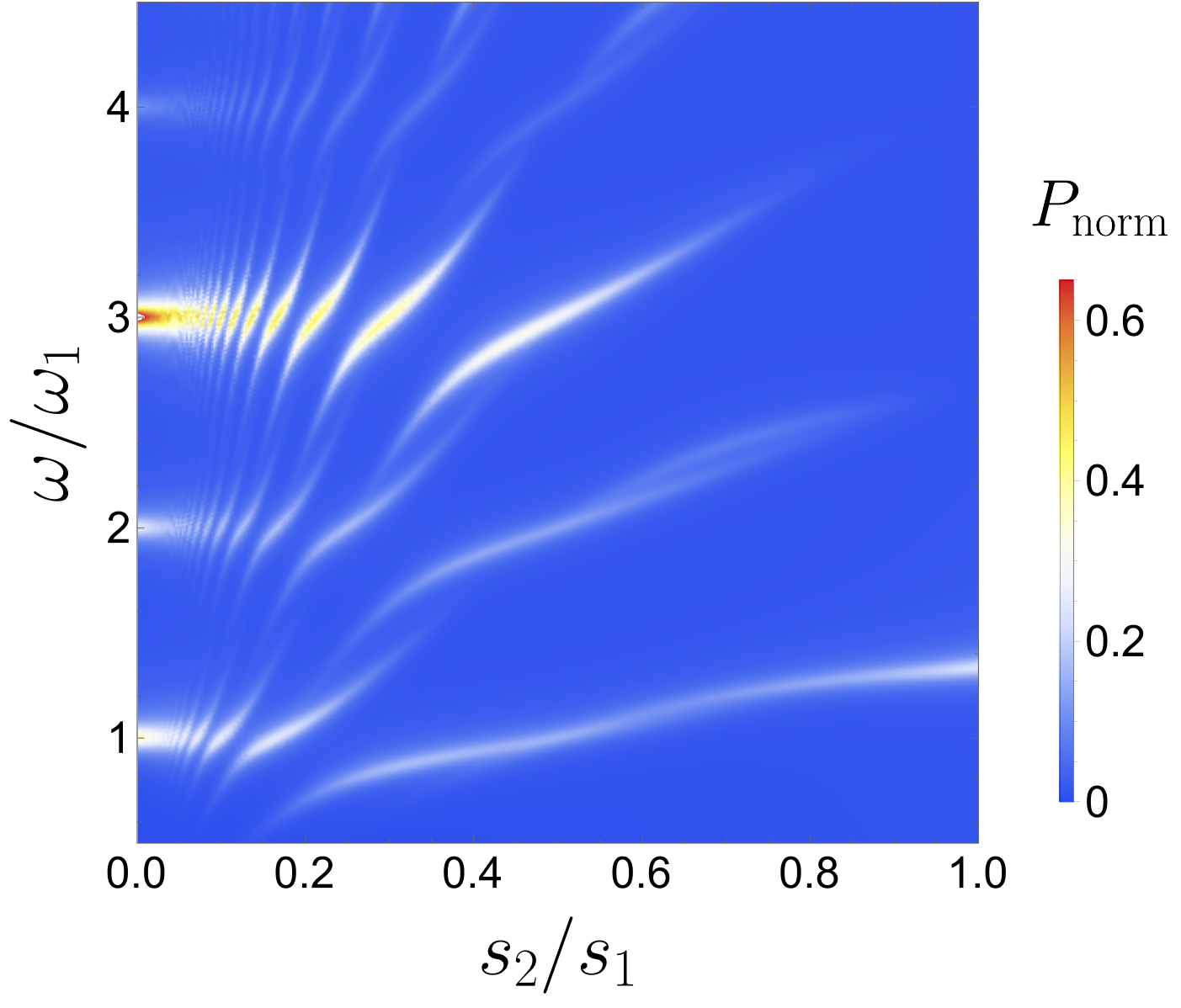}
\includegraphics[width=8.1 cm]{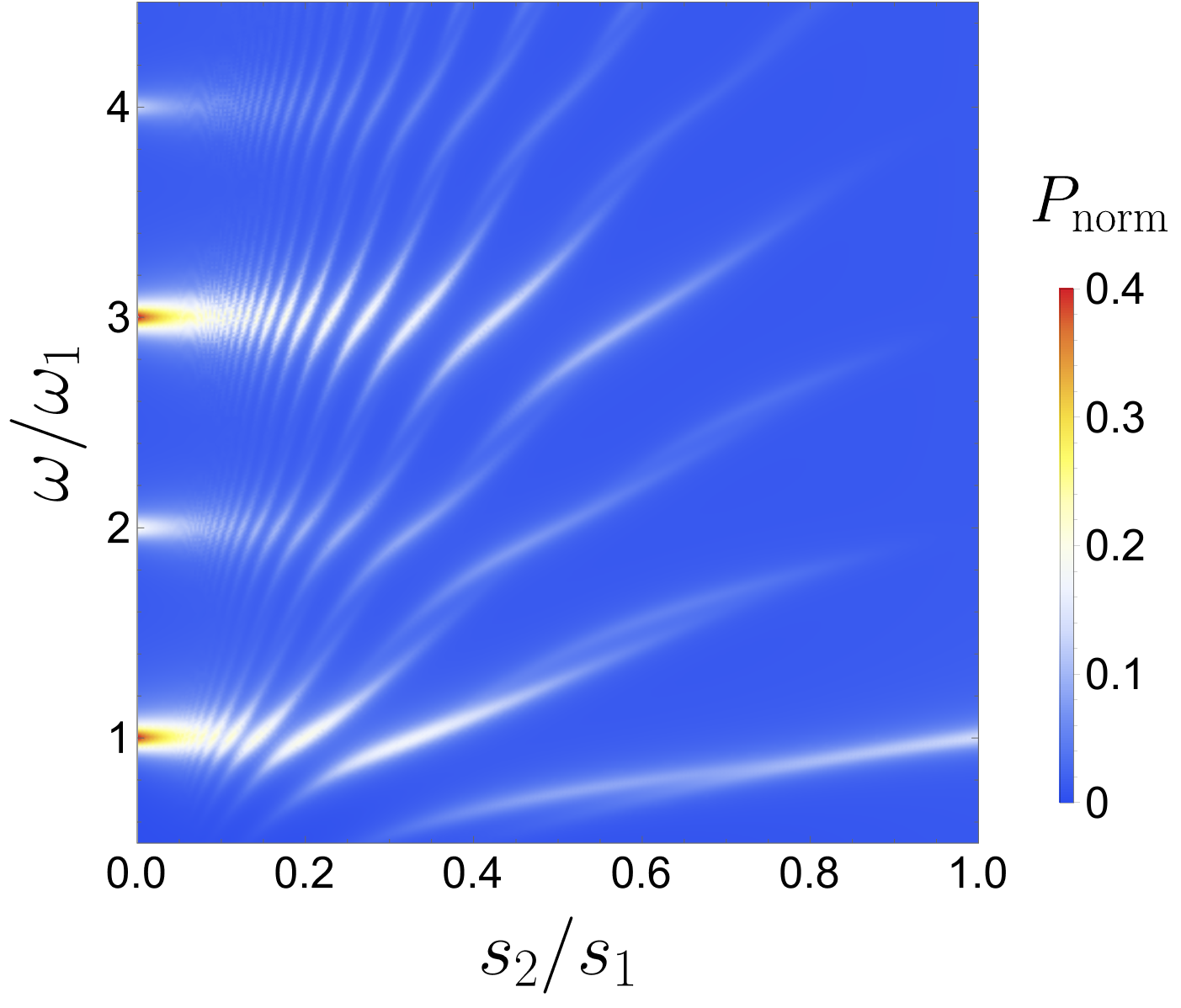}
\includegraphics[width=8.1 cm]{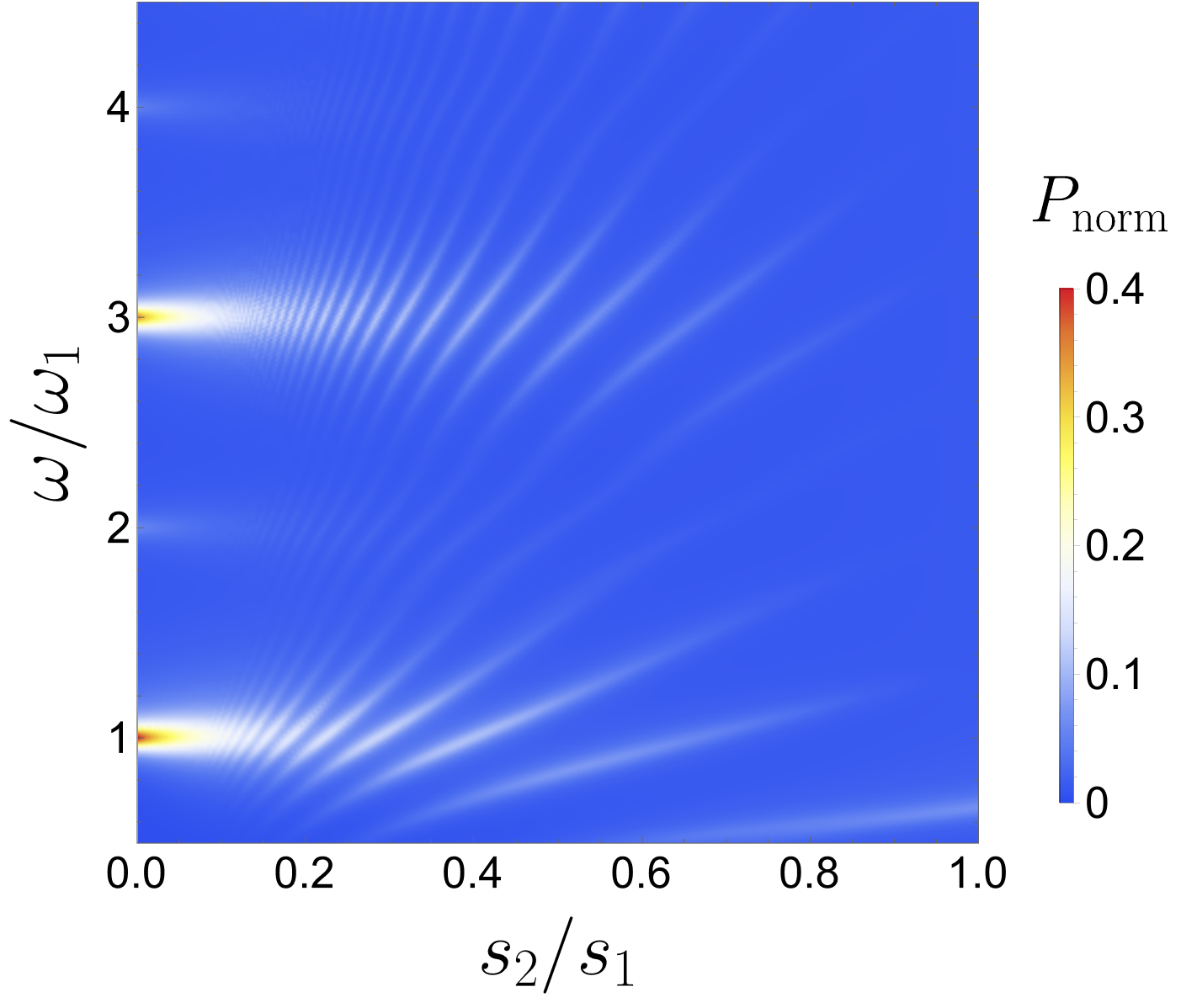}
\caption{ Heatmap of normalised dissipation $P_{\rm norm}$   for $\gamma = 0.1 \omega_1$, $\phi = 0$, $h = 0.5$ and different ratio $L_2/L_1$:   $L_2/L_1=0.5$ (upper panel), $L_2/L_1=1$ (central panel), and   $L_2/L_1=2$ (lower panel). Note difference in range of $P_{\rm norm}$. }
\label{Fig-s2_h_evolution}
\end{figure}

\subsection{Optical modulation by grating, $h \neq 0.$ }
\label{Sec-grating_modulation}
Next, we consider optical modulation of the incoming radiation described by Eq.~\eqref{Eq-grating_modulation} with a small but finite $h$ in the absence of the magnetic field.

Calculations are very similar to the ones  in Sec.\ref{HD}.
The values of radiation-induced   velocity $v_x^h$ and concentration $n^h$ are related to Eq.~\eqref{vxy}  as follows:
\begin{equation}
v_x^{\rm h} =v_x + h \frac{i F_0 \omega}{2 m s_\alpha^2(q_\alpha^2 - k^2)}  \cos{(k x+\phi)}, 
\end{equation}
\begin{equation}
\delta n^{\rm h} = \delta n - h \frac{F_0 k}{2 m s_\alpha^2(q_\alpha ^2 - k^2)}  \sin{(k x+\phi)},
\end{equation}
where $\alpha=1,2$ (here, we skip index $\alpha$ everywhere except terms, which are proportional to $h$).  Using these formulas and performing standard calculation described   in Sec.~\ref{HD}  and   Appendix \ref{AppHD}, one can find the unique solution   $v_x^{\rm h} (x) $ and $\delta n^{\rm h} (x)$ which is not growing for $|x| \to \infty.$   This solution should be  substituted into Eq.~\eqref{disintB}.  
  The resulting equation for the dissipation   turns out to be  too cumbersome for arbitrary $s_2/s_1$ and just as in the case $K \neq 0,$ we present plots calculated with the use of the exact formula for $P$ and also present analytical results for limiting cases of strong and weak coupling. 

In  Fig.\ref{Fig-s2_h_evolution} we present map  of normalized  dissipation, $$P_{\rm norm}=\frac{\omega^2 P}{\omega_1^2 P_0},$$  evaluated using exact analytical calculations  for different values  of $L_1/L_2.$  The normalization factor $\omega^2/\omega_1^2$  is used for better visualization   of the plots, which do not decay with $\omega$ as $1/\omega^2$  in contrast to $P.$ From this figure,  one can clearly see appearance of  the dark  modes at $\omega = 2 n \omega_1$ in the resonant regime.   The amplitudes of dark modes depend on ratio  $L_2/L_1.$ 

Next, we discuss simple limiting cases and demonstrate analytically that  dark modes indeed are sensitive to geometrical factor  $L_2/L_1$ as well as to  the phase $\phi$ responsible for asymmetry of the structure.

\subsubsection{Strong coupling,  $s_2 \to 0$}

In this case,  active regions are independent.   Optical  modulation on the grating period $L_1+L_2$  leads to inhomogeneous exciting field, which is the same  inside all  active  strips in contrast to the discussed above case $K \neq 0.$

Homogeneous component of the electric field excites bright modes  Eq.(\ref{Eq:Pstrong}), while the modulated component  excites dark modes with the amplitude  proportional to $h^2$ and also yields corrections to the amplitudes of  the bright modes:
\be
P= P_{\rm bright} + P_{\rm dark},
\ee
where
\begin{equation}
\begin{aligned}
&P_{\rm bright}= P_0 \sum_{m =0}^\infty \frac{\gamma^2 (B_m  + h C_m + h^2 D_{2 m + 1})}{\left[\omega- (2 m+ 1)\omega_1 \right]^2+\gamma^2/4}
\\
&P_{\rm dark}= P_0 \sum_{m =1}^\infty  \frac{ h^2 \gamma^2 D_{2 m}}{\left[\omega- 2 m \omega_1 \right]^2+\gamma^2/4}.
\end{aligned}
\label{dissGrating}
\end{equation}
Here $B_m$ given by Eq.~\eqref{Eq-Bm}, 
\begin{equation}
    C_m = \frac{2 s_1^2 \left[\cos{\phi} + \cos{(\phi+\frac{2 \pi L_1}{L_1+L_2}})\right]}{L_1 (L_1 + L_2) \left[\omega_1^2 (2 m +1)^2 - k^2 s_1^2\right]},
    \label{Eq-cm}
\end{equation}
and
\begin{equation}
\begin{aligned}
&D_n =  \frac{n^2 (k/k_1)}{ 4 \pi^2 \left[n^2 -  (k/k_1)^2 \right]^2 }
\\
&\times \left[\cos{\phi}- (-1)^n  \cos{\left( \phi + \frac{2 \pi L_1}{L_1+L_2}\right)} \right]^2.
\end{aligned}
\label{Eq-Dm}
\end{equation}
The coefficients $D_n$ with even indices   $n = 2 m $ yield amplitudes of   the dark modes (up to a factor $4 h^2 P_0$).   The terms with  odd indices  $D_{2 m + 1}  $  yield corrections to the bright modes. We see that for $n=2m$ 
\be
D_{2m} \propto  \cos\phi - \cos\left(\phi+ \frac{2 \pi L_1}{L_1+ L2 } \right).
\ee
For any $\phi,$ this equation  tends to zero both for $L_1 \gg L_2$ and for $L_1 \ll L_2.$ 

 Importantly,  $D_{2m}=0$ for $\phi=\phi_{\rm sym}$ [see Eq.~\eqref{phi-sym}], so that dark modes can be excited in the asymmetrical structures only.
Expressions for coefficients simplify for  $L_1 = L_2 = L/2$: 
\be
\begin{aligned}
&C_{ 0} = -\frac{\sin{\phi}}{2 \pi},\qquad C_{m \neq 0} = 0 , 
\\
&D_1 =\frac{\sin^2\phi}{16}, \qquad D_{2m+1}=0,~\text{for}~m>0 ,
\\
&D_{2m}= \frac{4 m^2 \cos^2\phi}{\pi^2 (4 m^2-1)^2},~\text{for}~m>0. 
\end{aligned}
\ee
Then, Eq.~\eqref{dissGrating} also simplifies
\be
\begin{aligned}
&P = \frac{P_0 \gamma^2}{(\omega-\omega_1)^2+\gamma^2} \left( \frac{1}{\pi^2}-\frac{h \sin{\phi}}{2 \pi} + \frac{h^2 \sin^2{\phi}}{16} \right)
\\
&+\frac{4}{9 \pi^2}\frac{P_0 \gamma^2}{(\omega-2 \omega_1)^2+\gamma^2}  h^2  \cos^2{\phi} + \widetilde{P},
\end{aligned}
\label{PtildeP}
\ee
where $\widetilde{P}$  includes high-harmonics of both dark and bright modes:
\be
\begin{aligned}
&\widetilde{P} = P_0  \sum_{n=1}^\infty  \frac{1 }{\pi^2 (1+2 n)^2} \frac{ \gamma^2}{[\omega-(2 n+1) \omega_1]^2+\gamma^2}
\\
&+P_0  h^2  \cos^2{\phi} \sum_{n=2}^\infty  \frac{4 n^2 }{\pi^2 (4 n^2 - 1)^2} \frac{ \gamma^2}{(\omega-2 n \omega_1)^2+\gamma^2}.
\end{aligned}
\label{tildeP}
\ee
Equation \eqref{PtildeP} contains contribution of the fundamental bright mode with small corrections induced by non-zero $h,$ contribution of the fundamental dark mode with the amplitude proportional to $h^2,$   and contribution of high harmonics  that rapidly decay with $n$  due to the numerical factors ${1}/{(1+2 n)^2}$  and  ${4 n^2}/{(4 n^2-1)^2}$ entering, respectively,   the sums in the upper  and lower lines  of 
 Eq.~\eqref{tildeP}.

 We see  that optical modulation by  grating field  leads to two main effects:  (i) the dark modes become visible in the spectrum and show resonances at $\omega/\omega_1 = 2, 4, 6, \dots$ with a small ($\propto h^2$)   amplitudes;  (ii) both bright and dark modes depend on the asymmetry parameter $\phi.$
 
\subsubsection{Weak coupling,  $s_1 -  s_2 \ll s_1.$}
Dissipation in   the homogeneous electron liquid, $s_1=s_2,$  excited by electric field  Eq.~\eqref{Eq-grating_modulation} is given by   sum of the Drude response  and plasmonic one coming, respectively, from the homogeneous and inhomogeneous components of the incoming radiation:   
\begin{equation}
    P_{s_2=s_1} =P_{\rm Drude}+  \frac{P_0}{2} \frac{h^2 \gamma^2 }{ (\omega-k^2 s_1^2/\omega )^2+\gamma^2 }.   \label{Eq-grating_plasmon}
\end{equation}

The plasmonic resonance  has standard asymmetric shape   as in the so-called  ``damped oscillator'' model (see discussion in Ref.~\cite{Boubanga-Tombet2020}). 

For small but finite $\delta s =s_1-s_2 \ll s_1,$ some corrections to Eq.~\eqref{Eq-grating_plasmon} arise. We  take these corrections into account using perturbation approach with respect to $\delta s$  up to the second order: 
\begin{eqnarray}
\delta P= P -  P_{s_2=s_1}=P_0 \sum_{n = 1}^\infty \frac{\gamma^2 A_n }{(\omega- \omega_n^{\rm weak})^2+\gamma^2/4}  \nonumber \\
+ \frac{P_0 \gamma^2 }{ (\omega-k^2 s_1^2/\omega )^2+\gamma^2 } \frac{8 h \, \delta s \, \omega_1^{\rm weak} \cos{(\phi+ \beta)} \sin{\beta}}{(L_1 + L_2) \omega^2}
\nonumber
\end{eqnarray}
where  $A_n \propto (\delta s)^2$ is  given by Eq.~\eqref{Eq-An} 
and  $\beta = \pi L_1/(L_1+L_2).$   We note here that similar to the Sec.\ref{Sec-weak_coupling} we do not take into account resonant frequency splitting that is linear to $\delta s$ when $L_1 \neq L_2$ [see Eq.~\eqref{Eq-delta_w}] or quadratic to $\delta s$ when $L_1 = L_2$ [see Eq.~\eqref{Eq-delta_w2}]. So latter equation does not include dark modes.   First line of this equation is proportional  to $\delta s^2.$ It represents  contribution  coming from homogeneous component of the external field and was obtained above [see Eq.~\eqref{Eq-P_weak}]. The second line is linear-in-$\delta s$ correction to the plasmonic resonance  from  Eq.~\eqref{Eq-grating_plasmon}.
 We notice that the latter correction $\propto h \delta s$ depends on $\phi$ and scales as $1/\omega^4$ at high frequency. 
Using defined $\phi_{\rm sym}$ we can rewrite latter equation:
\begin{eqnarray}
\delta P= P_0 \sum_{n = 1}^\infty \frac{\gamma^2 A_n }{(\omega- \omega_n^{\rm weak})^2+\gamma^2/4}  - \frac{P_0 \gamma^2 }{ (\omega-k^2 s_1^2/\omega )^2+\gamma^2 } \nonumber \\
\times \frac{8 h \, \delta s \, \omega_1^{\rm weak} (-1)^n \cos{(\phi-\phi_{\rm sym})} \sin{\beta}}{(L_1 + L_2) \omega^2}
\nonumber
\end{eqnarray}
Expression  for dissipation simplifies at
  $L_1 = L_2 = L/2$ reads:
 \begin{eqnarray}
   \delta P =P_0 \sum_{n = 1}^\infty \frac{\gamma^2 A_n }{(\omega- \omega_n^{\rm weak})^2+\gamma^2/4}  \nonumber \\
    + P_0 \frac{\gamma^2 }{ (\omega-k^2 s_1^2/\omega )^2+\gamma^2 } \left[ \frac{h^2}{2}  - \frac{8 h \delta s \omega_1 \sin{\phi} }{L \omega^2} \right].
    \label{Eq-Pweakh}
\end{eqnarray}

\section{Arbitrary grating field modulation}
In this section we assume arbitrary field modulation in he following form
\be
E= E_0\left[ 1+\sum_{n=1}^\infty h_n \cos{(k_n x+\phi_n}) \right],
\label{Eq-hn}
\ee
with $k_n = 2 \pi n/(L_1+L_2)$.
\subsection{Weak coupling}
In case of weak coupling, dissipation up to the second order in $\delta s$ and $h_n$ reads 
\be
\begin{aligned}
&P^{\rm weak}= P_{\rm Drude}+  P_0 \sum_{n = 1}^\infty \frac{\gamma^2 }{ (\omega-\omega_n^{\rm weak})^2+\gamma^2/4 }
\\
&\times \left[A_n +  \frac{4 \delta s h_n \cos{(\phi_n+ \pi a_n) \sin{\pi a_n }}  }{\omega_n^{\rm weak} (L_1+L_2)}  + \frac{h_n^2}{8}\right] ,
\end{aligned}
\label{Eq_Pweak_kn}
\ee
where  $A_n \propto (\delta s)^2$ is  given by Eq.~\eqref{Eq-An} and coeffitient $a_n = n L_1/(L_1+L_2).$ Terms $\propto h_n h_m$ are exactly zero due to averaging for $m \neq n$. Resonant frequency $\omega_n^{\rm weak}$ is given by Eq.~\eqref{Eq-w_weak}.

We see that that adding the  $h_n-$terms to the field expansion do not change the underlying physics, the response still has the same resonant frequencies with corrections to the resonant  amplitudes. 

\subsection{Strong coupling}
Using the same field expansion we can get for strong coupling in limit of isolates strip that response reads
\be
P^{\rm strong}= P_{\rm bright} + P_{\rm dark},
\ee
where
\begin{equation}
\begin{aligned}
&P_{\rm bright}= P_0 \sum_{m =0}^\infty \sum_{n =1}^\infty \frac{\gamma^2 ( B_m  + h_n \widetilde{C}_{mn} )}{\left[\omega- (2 m+ 1)\omega_1 \right]^2+\gamma^2/4}
\\
&+P_0 \sum_{m =0}^\infty \sum_{n =1}^\infty \sum_{l=1}^\infty  \frac{\gamma^2 h_n h_l \widetilde{D}_{mnl}^{\rm bright}}{\left[\omega- (2 m+ 1)\omega_1 \right]^2+\gamma^2/4}
\\
&P_{\rm dark}= \sum_{m =1}^\infty \sum_{n =1}^\infty \sum_{l=1}^\infty  \frac{\gamma^2 h_n h_l \widetilde{D}_{mnl}^{\rm dark}}{\left[\omega- 2 m \omega_1 \right]^2+\gamma^2/4}.
\end{aligned}
\label{Eq_strong_kn}
\end{equation}
Here $B_m$ is given by Eq.~\eqref{Eq-Bm},   
\begin{equation}
    \widetilde{C}_{mn} = \frac{2 L_1 \left[ \cos{\pi a_n} + \cos{(\phi_n+\pi a_n}) \right]}{(L_1 + L_2) \left[(2 m +1)^2 - 4 a_n^2 \right]},
    \label{Eq-cmn}
\end{equation}
\begin{equation}
\begin{aligned}
&\widetilde{D}_{mnl}^{\rm bright} = \frac{2 (1+2 m)^2 L_1}{\pi^2 (L_1 + L_2) }
\\
&\times \frac{\cos{\pi a_n}  \cos{\pi a_l} \cos{(\phi_n+\pi a_n}) \cos{(\phi_l+\pi a_l)}}{\left[(2 m +1)^2 - 4 a_n^2 \right]\left[(2 m +1)^2 - 4 a_l^2 \right]},
\end{aligned}
\label{Eq_D_bright}
\end{equation}
and
\begin{equation}
\begin{aligned}
&\widetilde{D}_{mnl}^{\rm dark} = \frac{m^2 L_1}{2 \pi^2 (L_1 + L_2) }
\\
&\times \frac{\sin{\pi a_n}  \sin{\pi a_l} \sin{(\phi_n+\pi a_n}) \sin{(\phi_l+\pi a_l)}}{\left[m^2- a_n^2 \right]\left[m^2 -a_l^2 \right]},
\end{aligned}
\label{Eq_D_dark}
\end{equation}
where $a_n = n L_1/(L_1+L_2).$ 

We see that similar to the weak coupling case, the  additional terms ($\propto h_n$) in the field expansion  do not change resonant frequencies and lead only to  corrections to the amplitudes of the resonant peaks.
\section{Nonzero magnetic field}
\label{Sec-MF}
Next, we  discuss effects induced by magnetic field.

\begin{figure*}
\includegraphics[width=0.315\textwidth]{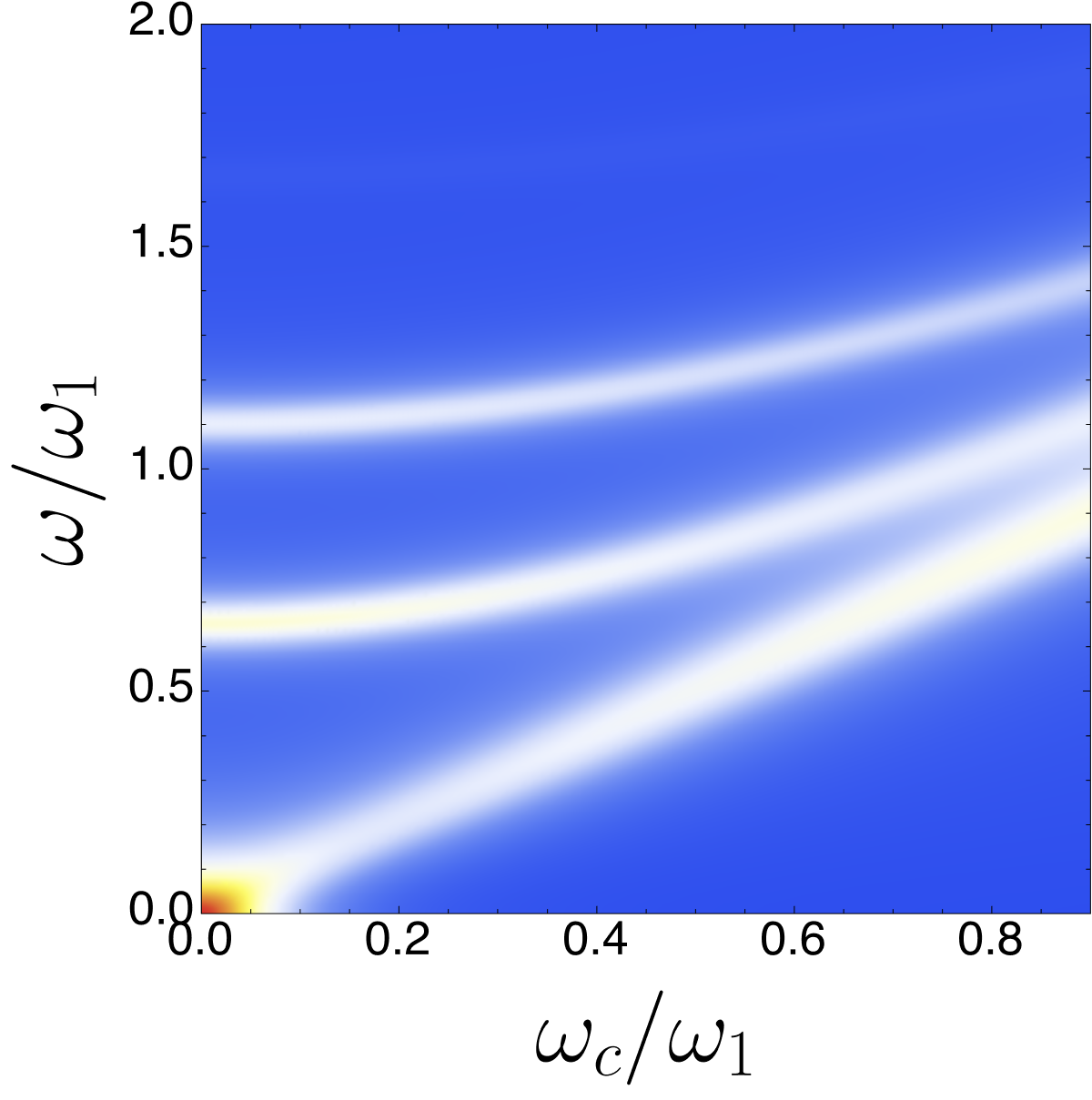}
\includegraphics[width=0.315\textwidth]{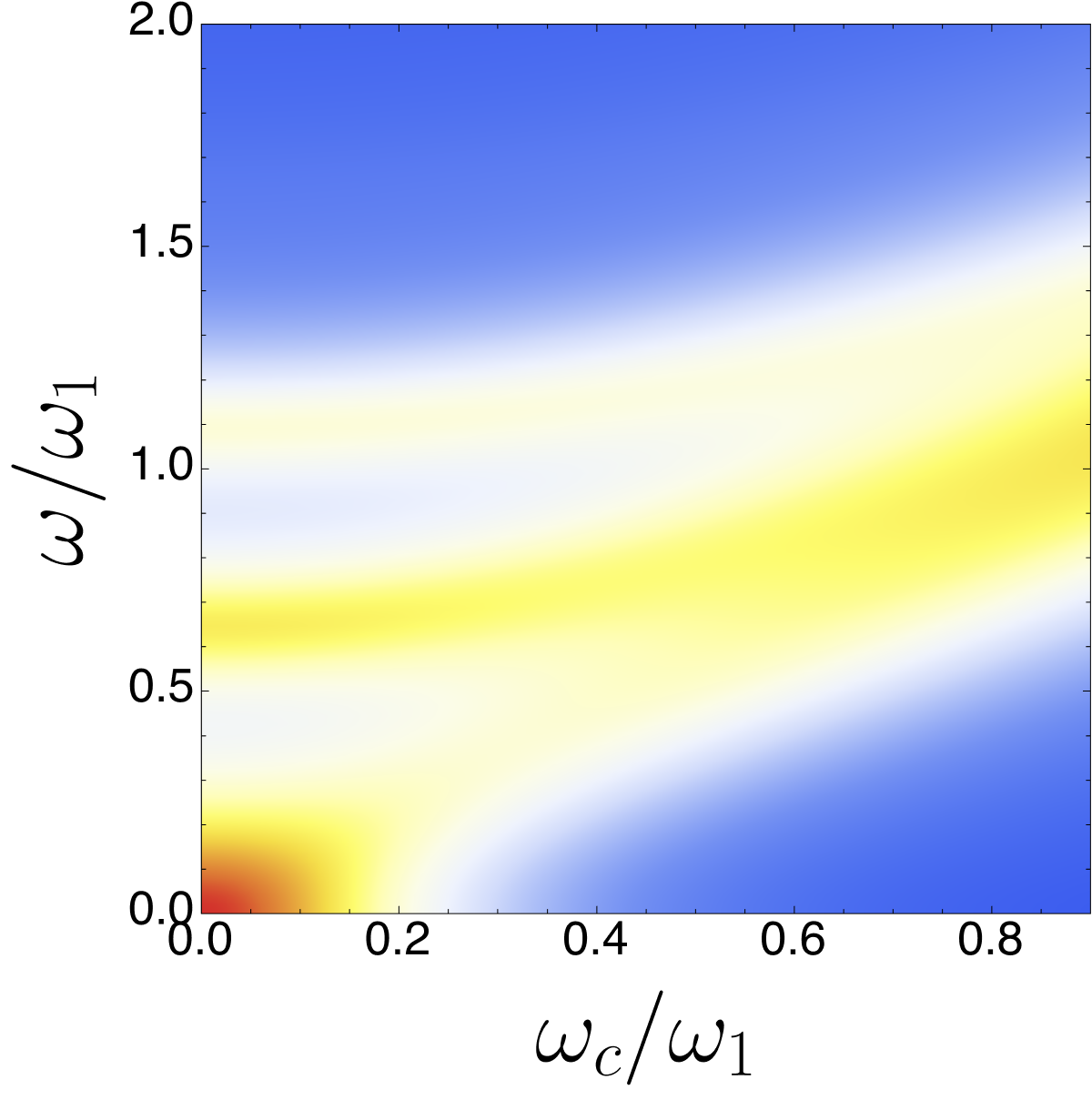}
\includegraphics[width=0.36\textwidth]{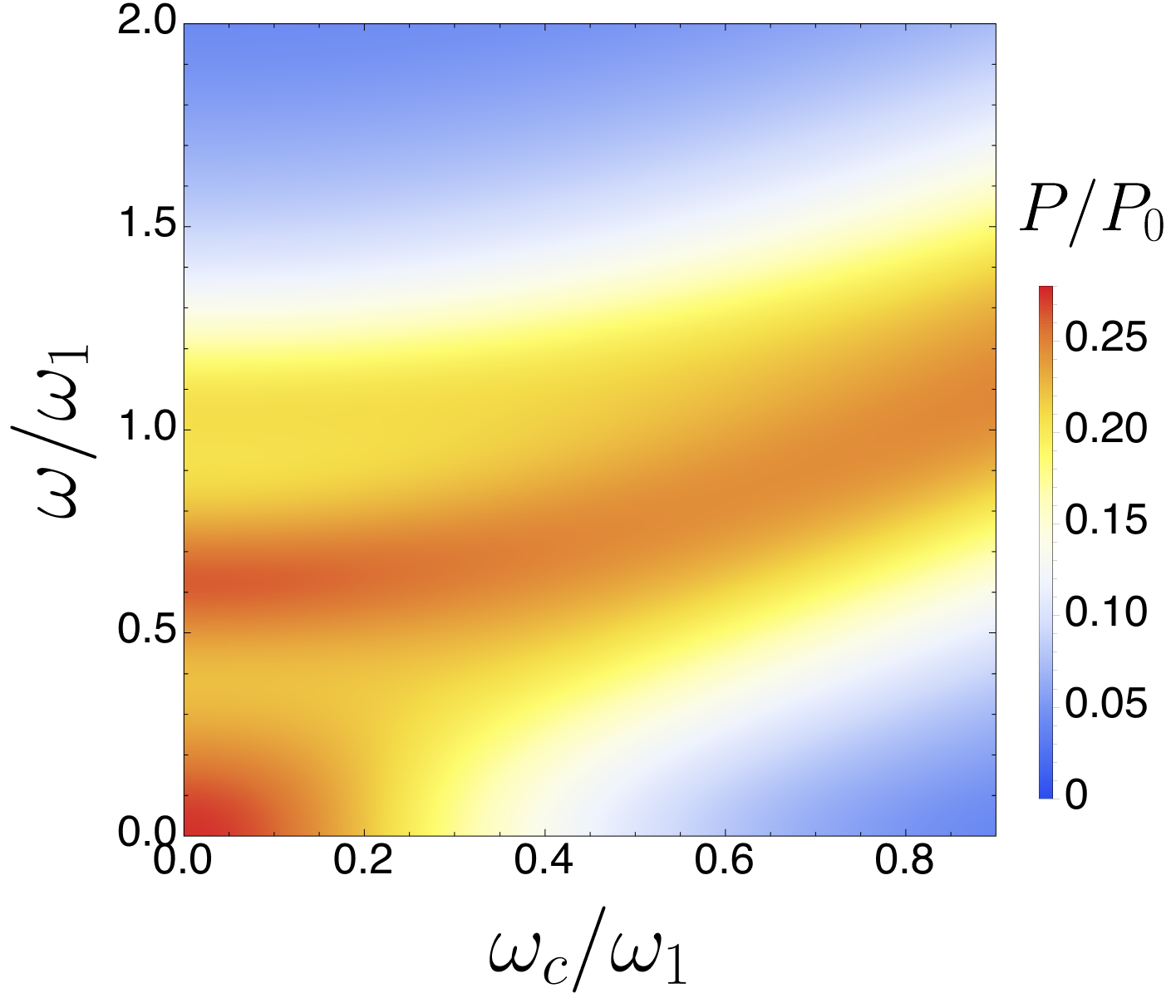}
\caption{ Heatmap of  dissipation in the $(\omega/\omega_1,\omega_c/\omega_1)$ plane for $s_2=0.4~ s_1,~L_2=L
_1,$   and  different values of $\gamma$:  $\gamma = 0.1~ \omega_1$(left panel),   $\gamma = 0.3~ \omega_1$ (central panel), and   $\gamma = 0.5 ~\omega_1$ (right panel).  Transition from super-resonant regime to resonant one with increasing of $\gamma$ and/or   $\omega_c$ is clearly seen.  }
\label{Fig-magnetic_evolution}
\end{figure*}
Heat map of the dissipation calculated with the   use of Eq.~\eqref{Eq-maindis} is shown in Fig.~\ref{Fig-magnetic_evolution} for fixed $\omega_{1,2}$ and different $\gamma.$ 
As seen, with increase of the magnetic field   the resonant frequencies increase, while the distance between neighboring resonances decrease. Therefore, magnetic field drive the system from super-resonant to resonant regime. Also,  the difference between  lowest plasmonic resonance  and cyclotron frequency decreases with $B,$ so that  for large magnetic field  plasmonic and cyclotron resonances overlap   due to the finite damping rate $\gamma$ and become indistinguishable.

Below, we discuss several  limiting cases, allowing for simple analytical description. We will limit ourselves to the case of absence of optical modulation, $h=0,~K=0.$

\subsection{Cyclotron resonance }
\label{Sec-cyclo}
\subsubsection{Homogeneous liquid}
 In the simplest case   $s_1=s_2,$  dissipation  reads:  \begin{equation} P= \gamma^2 P_0 \frac{|\gamma-i \omega|^2+\omega_c^2}{|(\gamma-i \omega)^2+\omega_c^2|^2}. \label{PC-general} \end{equation}
 For $\omega_c=0, $  Eq.~\eqref{PC-general} reproduces the Drude peak, $P_{\rm Drude},$  while for $\omega_c \gg \gamma$  and $|\omega -\omega_c| \ll \omega_c$ Eq.~\eqref{PC-general} shows  the  cyclotron resonance with the amplitude twice smaller than the Drude peak: 
\begin{equation} P_{\rm{C}}\approx  \frac{P_0}{2}  \frac{\gamma^2}{(\omega-\omega_c)^2+\gamma^2} = \frac{P_{\rm Drude}(\omega-\omega_c)} {2}.
\label{PC-CR}
\end{equation}

\subsubsection{Inhomogeneous liquid}
Let us consider now cyclotron resonance in the inhomogeneous liquid, where  
 $s_2 \neq s_1.$  Assuming $\gamma \ll \omega_c  \ll \omega_{1,2}, $  one can easily  find  from  Eq.~\eqref{Eq-maindis}  expression for cyclotron  resonance:
\begin{equation}
    P_{\rm cyclotron} =   \frac{P_0}{2} \frac{\gamma^2}{\gamma^2+(\omega-\omega_c)^2} \xi .
    \label{PC-Drude}
\end{equation}
This equation differs from Eq.~\eqref{PC-CR} only by the factor $\xi$ given by Eq.~\eqref{Eq-xi}.
It is worth noting that in the  strong coupling limit, $s_2 \to 0,$ i.e. for almost isolated active stripes,   cyclotron resonance is suppressed due to this factor: $\xi \propto (s_2/s_1)^2.$  

\subsection{
Strong coupling limit and resonant regime. }
In the  limit of very strong coupling we send  $s_2 \to 0,$ thus obtaining
\begin{equation}
    P = \left(1+\frac{\omega_c^2}{\omega^2+\gamma^2} \right)P_0 \frac{2 \gamma s_1}{\omega^2 (L_1+L_2) } {\rm Im} \left[ \frac{1}{s_1 \cot{q_1 L_1/2}}\right],
\end{equation}
where $q_1$ is determined by Eqs.~\eqref{Eq-wv_with_B} and \eqref{W-G} with $s=s_1.$
Depending on the relation between $\gamma$ and $\omega_1,$ this  equation allows to describe either resonant regime $\omega_2 \ll \gamma \ll \omega_1$ or non-resonant regime $\omega_1 \ll \gamma.$ Here, we restrict ourselves with a more interesting resonant regime.

The resonant frequencies, $\omega_n,$ are connected with zero-field equation \eqref{Eq-w_strong} in a standard way:      $\omega_n = \sqrt{(\omega_n^{\rm strong})^2+\omega_c^2}.$  Introducing $\delta \omega= \omega- \omega_n,$   in the resonant approximation ($|\delta \omega| \ll \omega_n$), we get
\begin{equation}
    P^{\rm strong} = P_0  \sum_{n=1}^\infty\frac{\gamma \gamma_n B_n}{\delta \omega^2+\gamma_n^2/4},
    \label{{Eq-PstrongB}}
\end{equation}
where damping rate is a function of resonant frequency $\gamma_n(\omega_n^{\rm strong}, \omega_c)$ with 
\begin{equation}
\gamma_n(\omega_n, \omega_c) = \gamma \frac{\omega_n^2 + 2 \omega_c^2}{\omega_n^2 + \omega_c^2} = \begin{cases}
    \gamma & \omega_c \ll \omega_n \\
    2 \gamma & \omega_c \gg \omega_n
		 \end{cases}.
\label{eq:gamma-n}
\end{equation}
Hence, the  effective damping rate increases in magnetic field. Analogous result was obtained for resonances in the isolated strip \cite{Zagorodnev2023}.

\subsection{Super-resonant regime and weak coupling limit}
In the super-resonant  regime, $\gamma$    is small as compared to other  parameters of  the system:
$\gamma \ll (\omega_{1,2}, \omega_c). $ 
The cyclotron resonance  in this regime was discussed above. Here, we focus on the effect of the magnetic field on the  sharp plasmonic super-resonances. 

One can show that Eq.~\eqref{eq-Pestimate}  is also valid for non-zero magnetic field, for $\omega> \omega_{\rm c},$ with the replacement
\begin{align}
&\omega_m \to \sqrt{\omega_m^2+ \omega_c^2},
\\
    & \mathcal L (\omega) \to  \frac{ \mathcal L \left(\sqrt{\omega^2-\omega_c^2} \right) (\omega^2- \omega_c^2 )^2}{\omega^4},
    \\
    &\Sigma_0 (\omega) \to  \frac{ \Sigma_0 \left(\sqrt{\omega^2-\omega_c^2} \right) (\omega^2- \omega_c^2 )^{5/2}}{ \omega^3 (\omega^2+ \omega_c^2 ) }.
\end{align}
Using these formulas, one can get simple analytical expression for the weak coupling case, $s_1- s_2 \ll s_1$:
\begin{equation}
    P = P_{\rm{C}} + P_0 \sum_{n=1}^\infty \frac{\gamma \gamma_n A_n}{\delta \omega^2+\gamma_n^2/4} 
    \label{Eq-P_weak_B}
\end{equation}
where $\delta \omega = \omega-\omega_n$, $\omega_n = \sqrt{\left(\omega_n^{\rm weak}\right)^2+\omega_c^2}, $ 
and $\gamma_n = \gamma_n (\omega_{ n}^{\rm weak}, \omega_c)$ [see   Eq.~\eqref{eq:gamma-n}].
The coefficient $A_n$ is given by Eq.~\eqref{Eq-An}.

\section{Discussion of experiment  }
\label{Sec-experiment}
\subsection{Comparison with recent experimental  results }

Next, we discuss  very recent experimental results  on measurements of transmission through grating gate GaN/AlGaN structure \cite{Sai2023} containing  about $10^3$  grating cells   (varying from sample to sample).   This structure   was illuminated  by  radiation with a frequency in a range of $0.5-3$ THz. Several fabricated grating gate structures with different  $L_1$ and $L_2$, both about 1 $\mu$m, were analysed.   

\begin{figure}[h!]
\includegraphics[width=0.5 \textwidth]{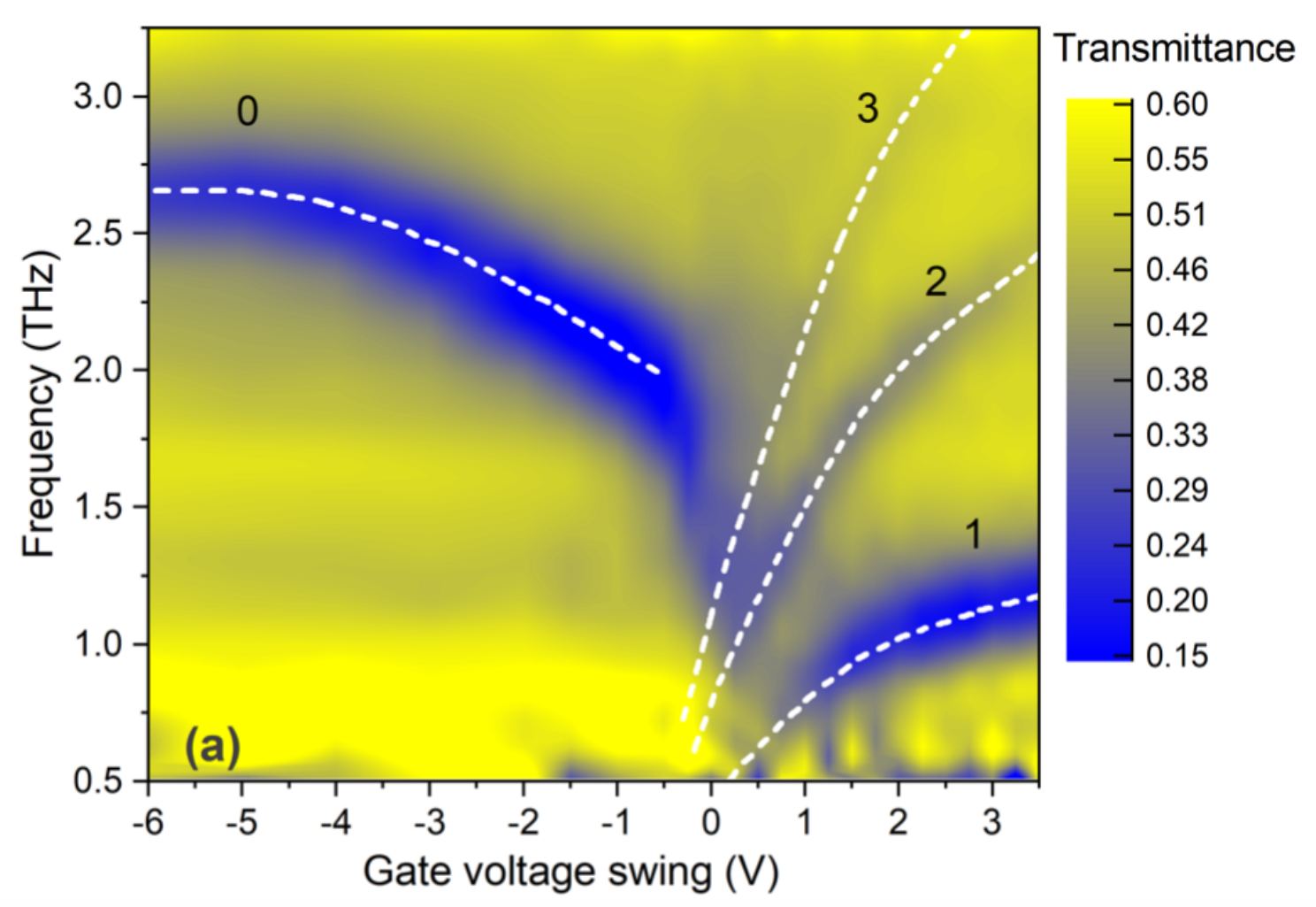}
\caption{Density plot of transmission coefficient adopted from \cite{Sai2023}. Here gate voltage swing is $V_0 = U_g^{(2)} - V_{\rm th}$.}
\label{Fig-experiment_density}
\end{figure}

\begin{figure}[h!]
\includegraphics[width=0.5 \textwidth]{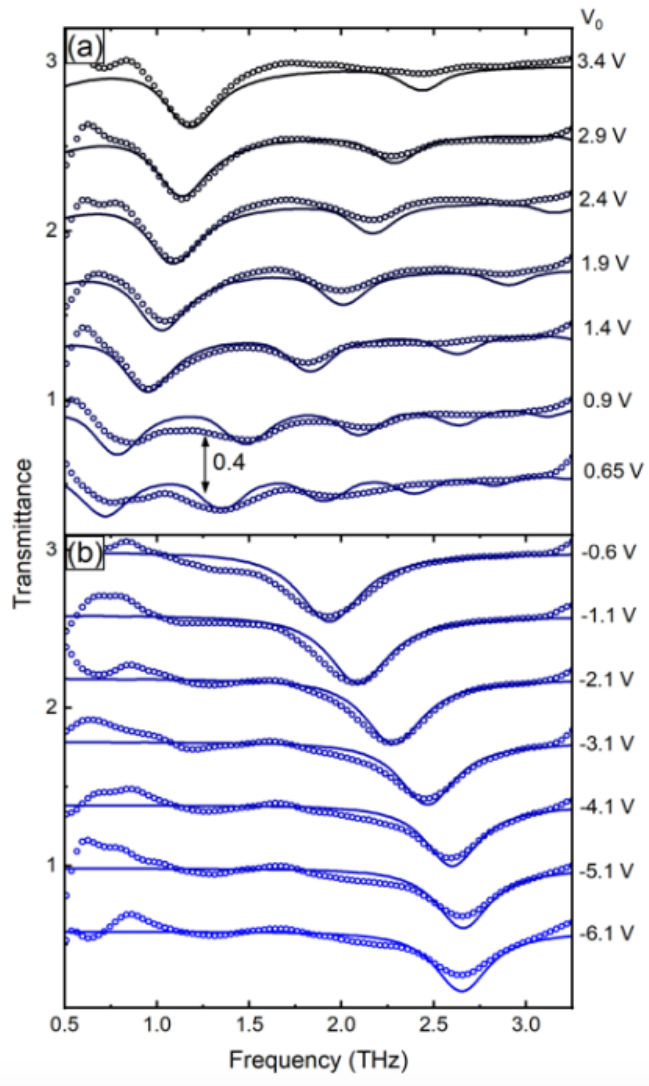}
\caption{Waterfall graph of transmittance adopted from \cite{Sai2023}. Points represent experimental values, solid lines - effective fitting.}
\label{Fig-experiment_waterfall}
\end{figure}

The experimental data presented in  Ref.~\cite{Sai2023}   are in very good agreement with our theory, although the experimental geometry was somewhat different from that discussed in this paper. Namely,   the active regions having  length  $L_1$ were ungated (while passive regions 2 were gated  just as  in our model). However, the  key physics of the problem is the same and one of the most interesting predictions of our theory---gate-tunable transition from super-resonant to resonant regime---was observed  in experiment  as clearly follows from comparison   of our Fig.~\ref{Fig-resfreq} with Fig.\ref{Fig-experiment_density} (Fig.~8 of  Ref.~\cite{Sai2023}).

  Let us discuss this question in more detail. The  transmission spectrum was measured in   Ref.~\cite{Sai2023}  for different values of gate voltage applied to passive regions.  Change of  this voltage  allowed tuning   the  electron concentration under grating strips.  
The weak coupling  super-resonant regime (called in Ref.~\cite{Sai2023} as delocalized phase) was realized  for gate voltages exceeding the threshold voltage, i.e. for positive gate voltage swing  ($V>0$ in notation of Ref.~\cite{Sai2023})   when the electron concentration in the passive region   was sufficiently high (this regime corresponds to red region of Fig.~\ref{Fig-regimes} with $s_1 \sim s_2$ and $\gamma \ll \omega_1$). Momentum relaxation time was estimated at \cite{Sai2023} as  $\tau = 0.85$ ps, which yielded sufficiently high quality factor to observe 
super-resonances. Specifically, three well resolved  resonances were seen  in the structure with $L_1=0.6~\mu$m and      $L_1=0.9~\mu$m  (see Fig. 8  of Ref.~\cite{Sai2023}).   
The strong coupling resonant regime    (called in Ref.~\cite{Sai2023} as localized phase)
 was realized  for  at negative gate voltage swing     (see  region $V <0$  in 
 Fig.~8  and panel  (b) in  Fig.~7  of Ref.~\cite{Sai2023}).  This regime corresponds to blue region in Fig.~\ref{Fig-regimes} with $s_2 \ll s_1$ and $\gamma \ll \omega_1.$
 
Hence, we arrive at  conclusion  that two regimes predicted by our theory  were clearly  seen in the experiment \cite{Sai2023}. Importantly, experiment  also shows sharp  gate-controlled transition between  these regimes in a full accordance with the theory. Indeed, with decrease of the gate voltage swing from positive   values the super-resonant frequencies decrease and move to zero (see Fig.~8 of Ref.~\cite{Sai2023}) in a good agreement with our theory (see, for example,  Fig.~\ref{Fig-resfreq}). With approaching $V$ to zero, the super-resonances merge and form a  single resonance  which survives  at $V<0.$ Such merging of super-resonances into single resonance   is in an excellent  agreement with our prediction  shown in Fig.~\ref{Fig-resfreq}  (see also Figs.~\ref{Fig-evolution} and \ref{Fig-evolution_gamma}).  
\color{black}

As seen from experimental data (see  region $V <0$  in 
Fig.~8, 9  in  Ref.~\cite{Sai2023}),    the resonant frequency for negative gate voltage swing depends on gate voltage, although  $V$ changes $s_2,$ not $s_1$.  The detailed discussion of this   dependence is out of scope of current work and will be presented elsewhere.   One of the possible mechanisms of such dependence was   proposed  at \cite{Sai2023}. It was suggested that effective width of ungated region $L_1$ decreases with increasing  $|V|$ at  $V<0,$ leading to  increase of $\omega_1,$ which is inversely proportional to  $L_1.$ 

Let us add two comments related to comparison of experimental data \cite{Sai2023} with our theory.   

According to the theory,  in  the strong coupling limit,  there should be a series of plasmonic resonances with frequencies $\omega_{1}, 3\omega_1, \ldots  $ [see  Eq.~\eqref{Eq-w_strong}]. Only lowest resonance with the frequency $\omega_1$ was observed  in experiment (although resonance at  $3\omega_1$ was predicted in numerical simulation presented in Ref.~\cite{Sai2023}). This can be explained by suppressing factor $1/(2 m +1)^2$ entering in the resonance at   $(2 m +1)\omega_1$ [see Eqs.~\eqref{Eq:Pstrong} and \eqref{Eq-Bm}].   
 
 The dark modes did not show up in  experiment \cite{Sai2023}, probably due to the   normal incidence of the  THz radiation.  We note, however, that the dark modes were observed in another experiment with similar geometry, where they 
were excited  by non-normal light  in the visible range of frequencies \cite{Hakala2017}.

\subsection{Theoretical predictions for experimental verification}
\label{Sec-tuning}
Transition from weak to strong coupling regime was observed in the experiment \cite{Sai2023} by depleting concentration is gated regions ($s_2 \to 0$) , so that ungated stripes    played the role of active regions and bright resonance with the frequency $\omega_1$ was observed. It would be very instructive to ``pump'' electrons in the same structure into gated region (instead of depleting it) so that $s_2$ would become larger that $s_1.$  
\begin{figure}[h!]
\includegraphics[width=0.5 \textwidth]{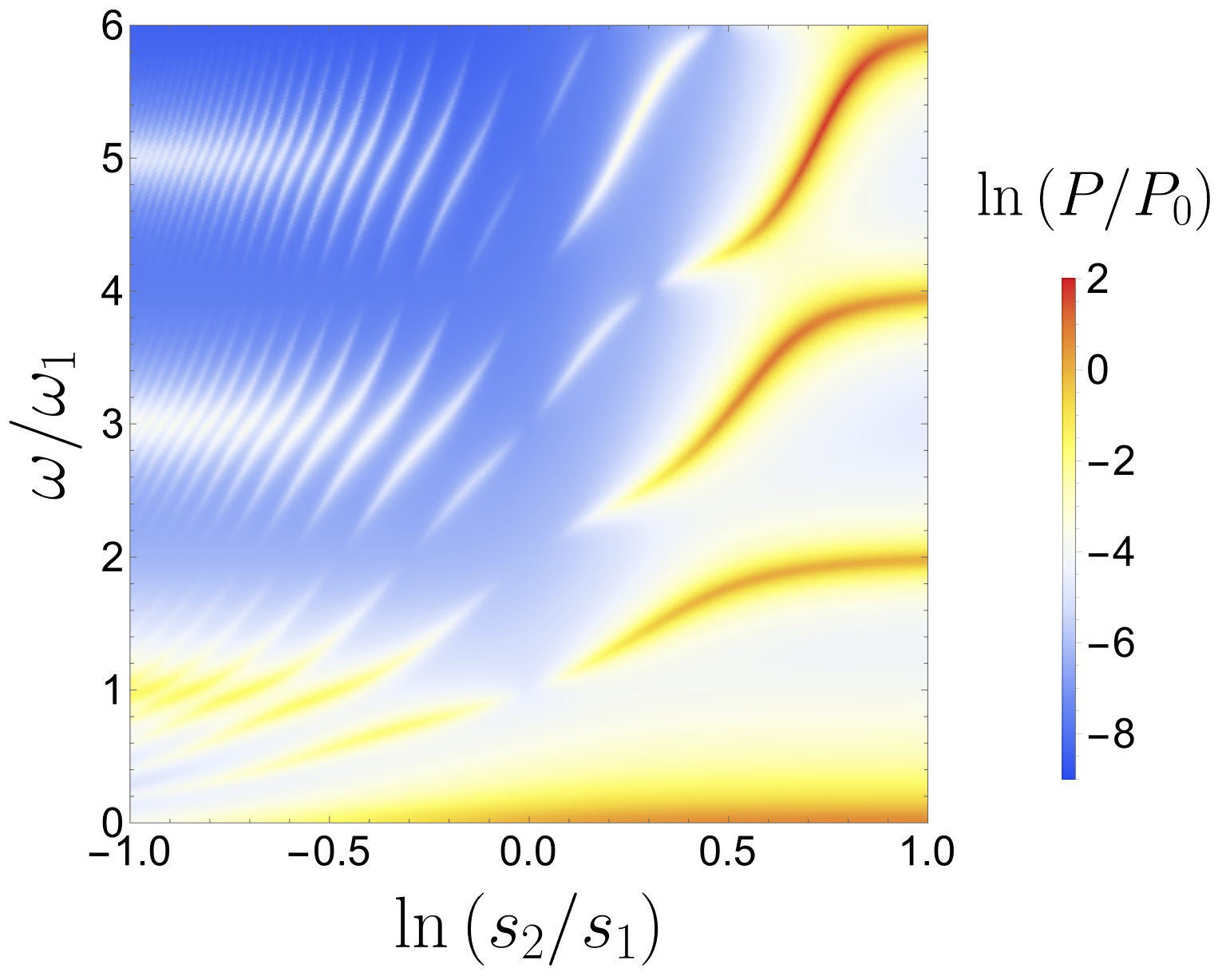}
\caption{   Transition from strong coupling resonant regime, $ \omega_2\ll\gamma\ll\omega_1$ (left part of the heatmap) to  strong coupling  inverted ($s_2>s_1$) super-resonant regime
$\gamma \ll \omega_1\ll\omega_2$ (right part of the heatmap) by ``pumping'' electrons (increasing $s_2$) into gated regions. Here $\gamma = \omega_1/5, L_2 = L_1.$ 
As seen,  resonances at frequencies $(2n+1)\omega_1$ for $s_2  \ll s_1$ transform into inverted super-resonances  at frequencies $2n \omega_1$ for $s_2 \gg s_1.$
}
\label{Fig-log_s_2}
\end{figure}

The transition from depleting to pumping is illustrated in  the Fig.~\ref{Fig-log_s_2}, where heat map is plotted for fixed $s_1$ and  in the wide interval of changing $s_2$: from very low concentration, corresponding to  resonant regime, $\omega_2 \ll \gamma \ll \omega_1,$ to very high concentration, corresponding to inverted (i.e. $s_2 \gg s_1$) super-resonant regime, $\gamma \ll \omega_1\ll \omega_2.$    The resonances are clearly seen both at small and large $s_2.$   However, the frequencies of resonances evolve from  odd resonant values, $(2n+1)\omega_1,$ for small $s_2 \ll s_1$ to even super-resonant values $2 n \omega_1 $ for inverted case, $s_2 \gg s_1.$ Physics behind this behavior is seen from Eq.~\eqref{Eq-Qbright}. 
For $s_2 \ll s_1$ resonant modes should be found from  condition  $\cos(\omega L_1/2 s_1)=0,$ while the super-resonant inverted modes for $s_2 \gg s_1$ obey $\sin(\omega L_1/2 s_1)=0.$     Experimental verification of  Fig.~\ref{Fig-log_s_2} would be a good control of the developed theory.

 \section{Conclusion}
To conclude,  we have developed  a theory of lateral  plasmonic crystal based on grating gate or double grating gate structures.   
We have shown that the spectrum of the crystal   is  controlled by     the voltages  on the  gates and, consequently,  can be tuned. 
We discussed transmission of THz radiation through  the crystal and  have found that only a part of plasmonic modes (bright  modes)   is seen in the transmission spectrum for the case of homogeneous excitation, while there also exist   dark modes  which  do not show up. 

We have analyzed  conditions for excitation of dark modes and found 
that they  can be excited  provided that the field of incoming radiation is inhomogeneous so that the whole system (radiation + multi-gate structure)  does not have inversion symmetry.

We   have  identified different modes of plasmonic oscillation:
(i) weak- and strong- coupling modes, the transition between which is tuned by the depth of concentration modulation; 
(ii) resonant and super-resonant modes, the transition between which is controlled by the momentum relaxation  rate and also by gate electrodes. 
Based on the developed theory, we have explained key features of very recent  experiment \cite{Sai2023}.

\section*{Acknowledgements}

We thank S.~L. Rumyantsev,  D.~V. Fateev,  and  V.~V. Popov for  useful discussions and comments. 
The work  was supported by the 
Russian Science Foundation under grant 24-62-00010.
The work of I.G. was also partially supported by the Theoretical Physics and Mathematics Advancement Foundation ``BASIS''.

\appendix

\section{Calculation of dissipation    for homogeneous excitation} \label{AppHD}
Here, we present some technical  details of calculations  in absence  of  external field modulation (i.e. for $h=0$ and $K=0$).  Linearizing the system of Eqs.~(\ref{Eq-Navier_Stokes}) and (\ref{Eq-continuity}) and searching  solution in the form  $\propto \exp(i q x -i\omega t)$     we get: 
\begin{equation}
\begin{aligned}
&(\gamma-i \omega) v_x -\omega_c v_y +i q s^2 \delta n =F_0/(2 m), 
\\
&\omega_c v_x +(\gamma-i \omega)v_y=0 , 
\\
&i q v_x -i \omega \delta n =0.
\end{aligned}
\label{hydrolin}
\end{equation}
Solving  these equations,  we  find Eqs.~\eqref{vxy} of the main text.  Using boundary conditions Eqs.~\eqref{Eq-BC} we find  transfer matrices $\hat T_{1,2}$ and functions $f_{1,2}$ entering Eqs.~\eqref{Eq-T-f}:
\be
\hat{T}_1 = \frac{s_1^2}{2 q_1 s_2^2}
\begin{pmatrix}
    e^{i q_1 L_1}(q_1 + q_2)  &  e^{-i q_1 L_1}(q_1 - q_2)  \\
    e^{i q_1 L_1}(q_1 - q_2) &  e^{-i q_1 L_1}(q_1 + q_2)
\end{pmatrix}
,
\ee
\be
\hat{T}_2 = \frac{s_2^2}{2 q_2 s_1^2}
\begin{pmatrix}
    e^{i q_2 L_2}(q_1 + q_2)  &  e^{-i q_2 L_2}(q_1 - q_2)  \\
    e^{i q_2 L_2}(q_1 - q_2) &  e^{-i q_2 L_2}(q_1 + q_2)
\end{pmatrix}
,
\ee
\be
f_1 =  \frac{i F_0 (q_1^2 - q_2^2)}{4 m q_1^2 q_2 s_2^2}, \, \, f_2 =  \frac{i F_0 (q_2^2 - q_1^2)}{4 m q_2^2 q_1 s_1^2}.
\ee
Next, using Eqs.~\eqref{Eq-T-f}, we find amplitudes $A$ and $B$ in both regions and substitute them into Eqs.~\eqref{vxy} thus obtaining solution for the whole PC.  In order to find dissipation, we first use the identity    
\begin{equation}
\begin{aligned}
&|\textbf{v}(x)|^2 = |v_x(x)|^2+|v_y(x)|^2 
\\
&=\left(1+ \frac{\omega_c^2}{\omega^2+\gamma^2} \right) |v_x(x)|^2,
\end{aligned}
\label{Eq-v_squared}
\end{equation}
 which directly follows from Eqs.~\eqref{hydrolin}. Then we rewrite Eq.~\eqref{Eq-diss0} as a sum  of dissipation over two regions:
\begin{equation}
\begin{aligned}
&P= \frac{2 m \gamma}{L_1+L_2}\!\!\left(1+\frac{\omega_{c}^2}{\omega^2+\gamma^2}\right)
\\
&\times \Bigr[ N_1 \!\!\int_{0}^{L_1} \!\!\!\!\!\! \left< |v_1(x, t)|^2 \right>_t \, dx +   N_2 \!\!\int_{L_1}^{L_1+L_2} \!\!\!\!\!\!\!\!\!\!\!\!\!\!\!  \left< | v_2(x, t)|^2\right>_t \, dx \Bigr] .
\end{aligned}
\label{disintB}
\end{equation}
and, finally, using analytical solutions for $v_1(x,t)$ and $v_2(x,t)$ we calculate integrals in Eq.~\eqref{disintB}.  After some  cumbersome but straightforward calculations we arrive at   Eq.~\eqref{Eq-maindis} of the main text.  

\section{Calculations for non-zero angle of radiation incidence.} 
\label{app:K-exp}
For $K \neq 0$ and  $\omega_c = 0$ we search  the solution for plasma wave velocity and concentration in the following form:

\begin{widetext}
    \begin{equation}
        \delta n_{M, \alpha}= \left[A_{M, \alpha} e^{i (q_\alpha-K) (x-L_M)}+B_{M, \alpha} e^{-i (q_\alpha+K) (x-L_M)} + \frac{F_0 K}{2 m s_\alpha^2 (q_\alpha^2-K^2)} \right]   \times e^{i K x -i \omega t} + c.c., 
    \end{equation}
    \begin{equation}
    v_{M, \alpha} = \left[\frac{\omega}{q_\alpha}  A_{M, \alpha} e^{i (q_\alpha-K) (x -L_M)} -  \frac{\omega}{q_\alpha} B_{M, \alpha} e^{-i (q_\alpha+K) (x-L_M)} + \frac{i F_0 \omega }{2 m s_\alpha^2(q_\alpha^2-K^2)}   \right]   \times e^{i K x - i \omega t} + c.c.
    \end{equation}
\end{widetext}
where $M$ and $\alpha = 1, 2$ numerate cells and regions within a cell, respectively, and $L_M=M (L_1+L_2).$ 
We search for solutions that are finite at $|M| \to \infty $  for the coefficients $A_{M, \alpha}$ and $B_{M, \alpha}$.
It is worth noting that these finite at large $|M|$  solutions for concentration and velocity change from cell to cell due to the factor $\times e^{i K x}$. At the same time, direct calculations show that   the dissipation $P$ is the same for all cells.

\bibliography{main}

\end{document}